\documentclass[%
reprint,
amsmath,amssymb,
floatfix,
showkeys
]{revtex4-2}

\usepackage[utf8]{inputenc}
\usepackage{url}
\usepackage{graphicx}
\usepackage{braket}
\usepackage{float}
\usepackage{array}
\usepackage{geometry}
\usepackage{dcolumn}
\usepackage{bm}
\usepackage{textcomp}
\usepackage{gensymb}
\usepackage{amsmath}
\usepackage{caption}
\usepackage{subcaption}
\usepackage{physics}
\usepackage{amssymb}
\usepackage{dsfont}
\usepackage{epsfig,amsmath,amsfonts}
\usepackage[hidelinks]{hyperref}
\usepackage{multirow}
\usepackage{rotating}
\usepackage{color}
\usepackage{fancyhdr}
\usepackage{lipsum,tikz}
\usepackage{tabularx, booktabs}
\usepackage{etoolbox}
\usepackage{upgreek}

\begin{document}

\title{Interference aided finite resonant response in an undamped forced oscillator\vspace{-0.3em}}
\author{Shihabul Haque}
\email{Corresponding author: ug2020sh2371@iacs.res.in (alt: shihabul1312@gmail.com)}
\affiliation{School of Physical Sciences, Indian Association for the Cultivation of Science, Kolkata, India \vspace{-0.5em}}
\author{Jayanta K. Bhattacharjee}
\email{jayanta.bhattacharjee@gmail.com}
\affiliation{Department of Physics, Indian Institute of Technology, Kanpur, India}

\date{\footnotesize{\today}}    

\begin{abstract}
We apply perturbative techniques to a driven undamped sinusoidal oscillator at resonance. The angular displacement, $\theta$, obeys the dynamics $\Ddot{\theta}+
\omega^{2}\sin\theta = H\cos\omega t$. The linearized approximation gives a divergent response (at long times) but the nonlinear terms make the response finite.
We address the nonlinearity-induced finiteness in two ways by separately treating the short and long time scales. At long times, we use the traditional perturbative
techniques to extract two drive dependent behaviours - one, the amplitude of oscillation scales as $(H/\omega^{2})^{1/3}$ and, two, the time period of the slow 
mode varies as $(H/\omega^{2})^{-2/3}$. For the early time behaviour, on the other hand, we devise an alternate perturbative expansion where the successive terms
get larger with the order of evaluation but have alternating signs. The alternating signs (phase differences) between these terms leads to adestructive interference 
like effect. A careful consideration of this destructive interference like effect between successive terms leads to a finite response which describes the initial 
behaviour of the amplitude of the response reasonably correctly. We further note that for larger drive values, the system seems to undergo a first order transitional 
behaviour with a sudden jump in the largest Lyapunov exponent

\end{abstract}


\maketitle

\section{Introduction}\label{sect1}
Perturbation theory near resonance generally implies computation of small changes in the resonance frequency or calculations of small changes in the response 
functions at resonance because of small deformations in the assumed ideal shape of the resonating object. A detailed study for deformed dielectric spheres has been 
carried out recently in \cite{aiello} and \cite{aiello2}. Similar examples are found in optical setups \cite{foreman} and in Helium drops \cite{childress}. Nonlinear 
vibrations in strings has been studied in \cite{elliot}. \\
In this work, we look at an undamped nonlinear system at resonance (analysed briefly in a formal manner in \cite{Jean}) – we consider an undamped forced pendulum in 
a uniform gravitational field with $\theta$ being the angle with the vertical. Related analyses of this system include \cite{Yorke} (who, however, consider damping 
as well which we do not) and \cite{harish} (who briefly consider this system and derive some interesting numerical results in the context of diffusion). The equation 
of motion for this pendulum takes the form, 
\begin{figure}[H]
	\includegraphics[width = \columnwidth, height = 50mm]{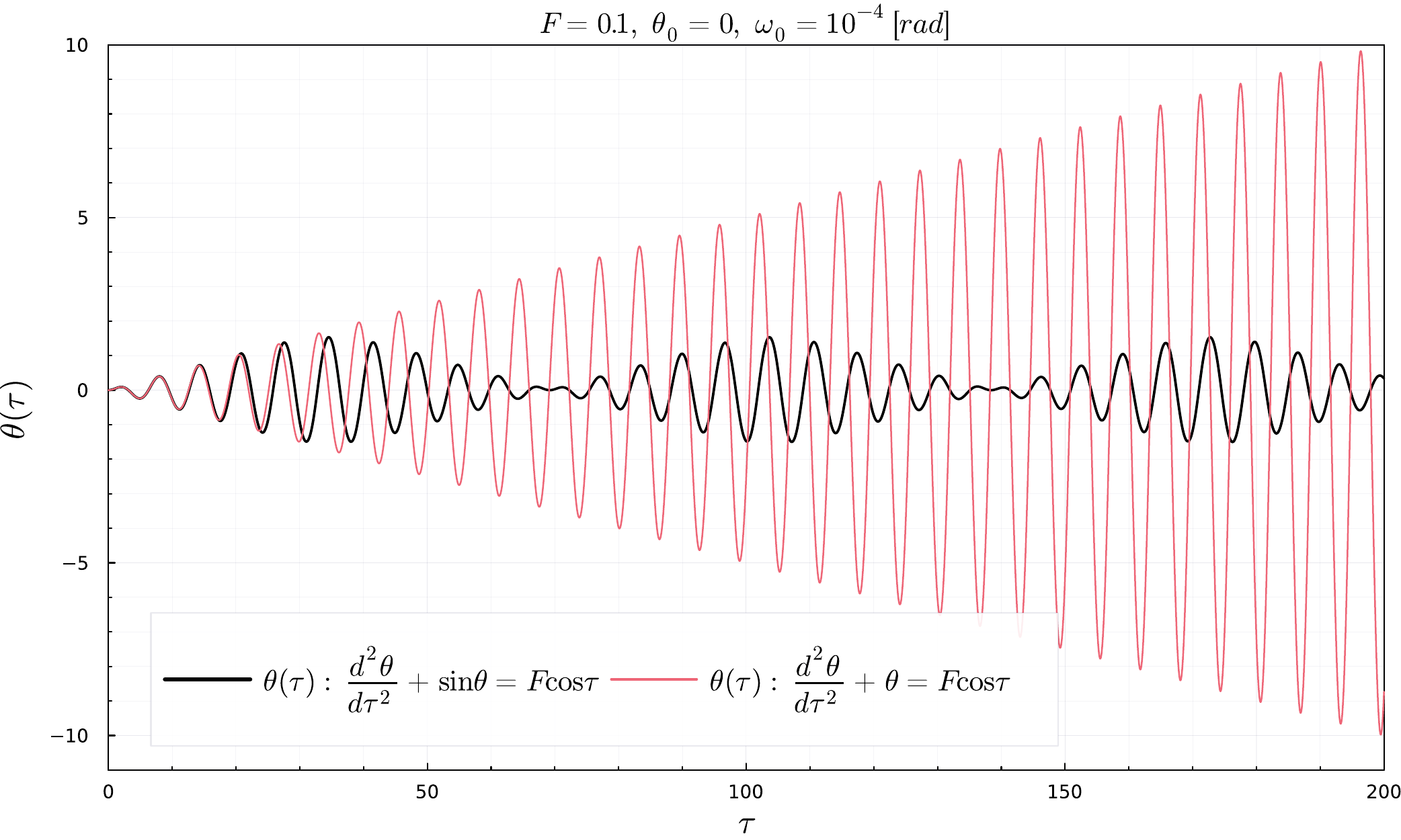}
	\caption{The forced oscillator with and without the linear approximation}
  \label{figdis}
\end{figure}
\begin{equation}\label{or0}
    \Ddot{\theta} \ + \ \omega^{2}\sin{\theta} \ = \ H\cos{\omega t}
\end{equation}
In the linear approximation, the system is described by the equation, $\Ddot{x} \ + \ \omega^{2}x \ = \ H\cos{\omega t}$, with the particular solution, 
$\frac{H}{2\omega}t \sin{\omega t}$. This is a diverging solution - from the red curve in Fig. (\ref{figdis}), we see that it differs significantly from the 
finiteness of the total solution describing eqn. \eqref{or0} (the plot reflects the re-scaled problem with $\tau = \omega t$, $H = \omega^{2}F$ being dimensionless; 
$\omega_{0} \equiv \Dot{\theta}(0)$ is used through the text). \\
The first order version of this kind of dynamics, $\Dot{\theta} + \sin\theta = r\cos t$, was studied in \cite{Adler} a decade ago. But, as 
far as we know, the corresponding second order dynamics has never been investigated in detail. The reason probably lies in the fact that eqn. \eqref{or0}, being a 
non-autonomous second order dynamical system, is susceptible to chaotic behaviour and the expectations could have been that interesting oscillatory behaviour would 
not be on the cards. We have found, on the contrary, that the dynamics of $\theta (t)$ become quite complicated when non-chaotic and conventional perturbative 
techniques give only a qualitative picture, failing to provide the magnitude of the oscillation amplitude correctly. We show the comparison in section \ref{sect2}, 
noting the discrepancies, after introducing a different kind perturbative expansion in section \ref{sect3} which leads to a series of diverging terms but at the 
$n^{th}$ order gives a clear description of the real time series up to a time, $\tau_{n}$, which increases with $n$. This is a non trivial agreement because the 
oscillation amplitudes at the $n^{th}$ order increase in magnitude with increasing $n$ but differing phases at various orders allows the perturbative solution to 
reasonably approximate the exact answer for the interval $\tau_{n}$ - at any given order, for $t > \tau_{n}$, the discrepancy between the actual and approximate 
dynamics is large, but unlike the traditional techniques where the amplitudes never match, this allows a correct description over an ever increasing time interval. 
Boundedness of such nonlinear differential equations and resonant behaviour have been studied rigorously in \cite{lazer} and \cite{alonso}.\\
Section \ref{sect3} details our perturbative approach where we explain the finiteness of the oscillations as a result of varying phases and destructive interference. 
The traditional approaches to eqn. \eqref{or0} are described in detail in section \ref{sect2} where we show that the maximum amplitude of the oscillator scales with 
the driving term magnitude as $F^{1/3}$. We also show that the slower timescale evident from the behaviour depicted in Fig. \ref{figdis} is driving dependent - 
specifically scaling as $F^{-2/3}$. We use both the Linstedt - Poincaré and the Krylov - Bogoliubov techniques and find that there is a considerable overlap of 
results over a large range indicating the robustness of the perturbative technique. We further look at some interesting features of the system in section \ref{sect4}, 
pointing out an interesting transitional behaviour in the system with respect to the driving term - specifically, we show that there is a jump in the maximum Lyapunov 
exponent of the system as $F > 0.33$ indicative of a first order transition from the ordered to the chaotic state. A brief discussion is given in section \ref{sect5}.

\section{Interference aided regularization}\label{sect3}
\begin{figure}[H]
    \centering
        \includegraphics[width = \columnwidth]{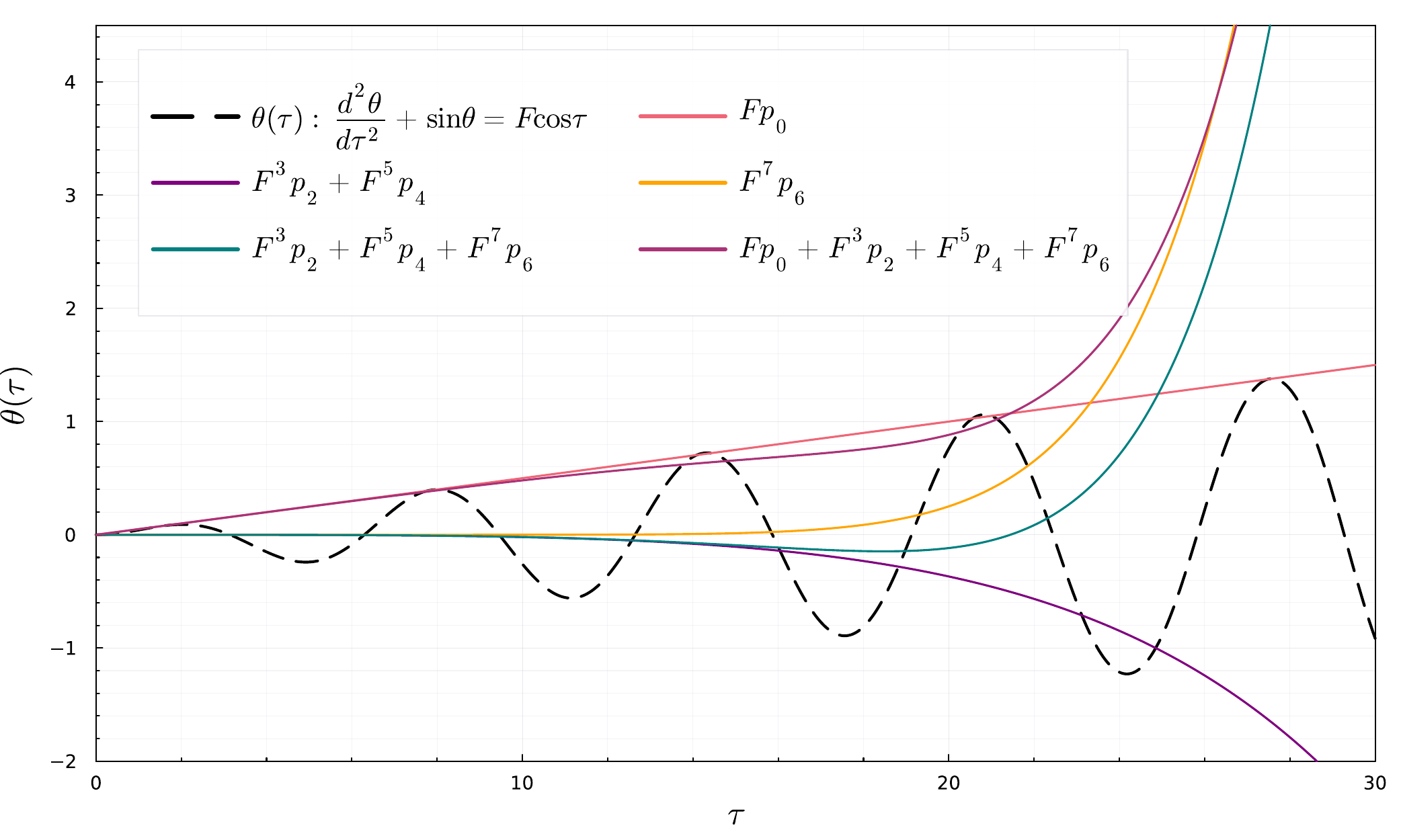}
        \caption{The forced oscillator polynomials of various order for $F = 0.1$}
        \label{figu1}
\end{figure}

In this section, we propose a new perturbative approach to explain the finiteness of the response displayed in Fig. \ref{figdis}. The original solution corresponds 
to the linear approximation to \eqref{or0}; therefore, we begin by rescaling the problem to make $H = \omega^{2}F$ dimensionless and write,
\begin{equation}\label{mid}
    \Ddot{\theta} \ + \ \omega^{2}\sin{\theta} \ = \ \omega^{2}F\cos{\omega t}
\end{equation}
This enables us to use $F$ as the regulator in the problem; the original solution to the problem is given by the solution of the equation,
\begin{equation} \Ddot{\theta}_{0} \ + \ \omega^{2}\theta_{0} \ = \ \omega^{2}F\cos{\omega t}\end{equation}
We introduce the small correction to the above as,
\begin{equation} \Lambda \ = \ \omega^{2}(\sin{\theta} \ - \ \theta)\end{equation}
In order for this to be small (and for perturbation theory to be applicable), we require that $\theta$ be small and in order to achieve this we expand the variable 
in a specific manner assuming that $F$ is very small. From \eqref{mid}, after scaling $\theta$ as $F\theta$,
\begin{equation}
    \Ddot{\theta} \ + \ \omega^{2}\theta \ + \ \Lambda \ = \  \Ddot{\theta} \ + \ \omega^{2}\sin{\theta} \ = \omega^{2}F\cos{\omega t}
\end{equation}
Now, we expand $\theta$ as $F\theta_{0} + F^{2}\theta_{1} + F^{3}\theta_{2} + ... $, yielding us in the zeroth order (or $\mathcal{O}(F)$) with $\sin{\theta} 
\approx \theta = F\theta_{0}$,
\begin{equation} \Ddot{\theta}_{0} \ + \ \omega^{2}\theta_{0} \ = \ \omega^{2}\cos{\omega t}\end{equation}
This is the desired leading order equation that gives us the original solution to the problem to which the corrected solution will be added. The complete solution 
to this equation is, 
\begin{equation}\nonumber
    \theta_{0}(t) \ = \ A\cos{\omega t} + (B + \frac{\omega t}{2})\sin{\omega t}
\end{equation}
\begin{equation}\label{zero}
     \ = \ (\alpha + \frac{\omega t}{2})\sin{\omega t}
\end{equation}
Here we impose the initial conditions $\theta_{0}(t = 0) = 0; \ \dot{\theta}_{0}(t = 0) = \omega\alpha$.\\
In the next higher order, $\mathcal{O}(F^{2})$, we match powers of $F$ and end up with, 
\begin{equation}
    \Ddot{\theta}_{1} \ + \ \omega^{2}\theta_{1} \ = \ 0 
\end{equation}
The solutions to this are very well known; however, we equal the initial velocity and displacements (in this order) to 0, effectively giving us no contribution from 
this order. In the next order, $\mathcal{O}(F^{3})$, we get a more interesting equation,
\begin{equation}\nonumber
    \Ddot{\theta}_{2} \ + \ \omega^{2}\theta_{2} \ = \ \frac{\omega^{2}}{6}\theta_{0}^{3} 
\end{equation}
\begin{equation}\label{three}
\ = \ \frac{\omega^{2}}{24}\Big(\alpha + \frac{\omega t}{2}\Big)^{3}(3\sin{\omega t} - \sin{3\omega t}) 
\end{equation}
This is the equation of a forced harmonic oscillator with the forcing term proportional to the zeroth order solution. As in \cite{Lai, olsson}, we neglect the 
higher frequency term in this approximation and keep the largest contribution from the cubic term. Scaling $t$ as $\tau = \omega t$ and using the double dot 
notation to denote a derivative with respect to the latter, we get, 
\begin{equation}
    \frac{d^{2}\theta_{2}}{d\tau^{2}} \ + \ \theta_{2} \ = \ \Ddot{\theta}_{2} \ + \ \theta_{2} \ = \ \frac{1}{64}\tau^{3}\sin{\tau} \ = \ G(\tau)
\end{equation}
The particular solution to the above forced oscillator is given by, 
\begin{equation}\nonumber \theta_{2}(\tau) \ = \ \end{equation}
\begin{equation}\nonumber \sin{\tau}\int_{0}^{\tau}G(\xi)\cos{\xi} \,d\xi -  \cos{\tau}\int_{0}^{\tau}G(\xi)\sin{\xi} \,d\xi \end{equation}
\begin{equation}
\ = \ I_{1}\sin\tau  - I_{2}\cos\tau \end{equation}
Here we have chosen the homogenous solutions to be $\sin\tau, \cos\tau$ such that the Wronskian of the differential equation is $-1$. Then, we have,
\begin{equation} I_{1} \ = \ \frac{1}{512}\Big((3\tau-2\tau^{3})\cos2\tau + \frac{3}{2}(2\tau^{2} - 1)\sin2\tau\Big) \end{equation} 
\begin{equation} I_{2} \ = \ \frac{1}{512}\Big(\tau^{4} + (3\tau-2\tau^{3})\sin2\tau + (\frac{3}{2} - 3\tau^{2})\cos2\tau - \frac{3}{2} \Big) \end{equation} 
Taking the largest term, we get, $$\theta_{2}(\tau) \sim -\frac{1}{512}\tau^{4}\cos\tau$$ The negative sign tells us that in this order the amplitude drops rapidly, 
in constrast to the leading order where the signs to the solution were positve denoting a growing contribution (more accurately, this is a phase change). In a 
similar fashion, we move to the next order, $\mathcal{O}(F^{4})$, 
\begin{equation} \Ddot{\theta}_{3} \ + \ \omega^{2}\theta_{3} \ = \ \frac{1}{2}\theta_{0}^{2}\theta_{1} \end{equation}
\begin{figure*}
	\includegraphics[width = \textwidth, height = 85mm]{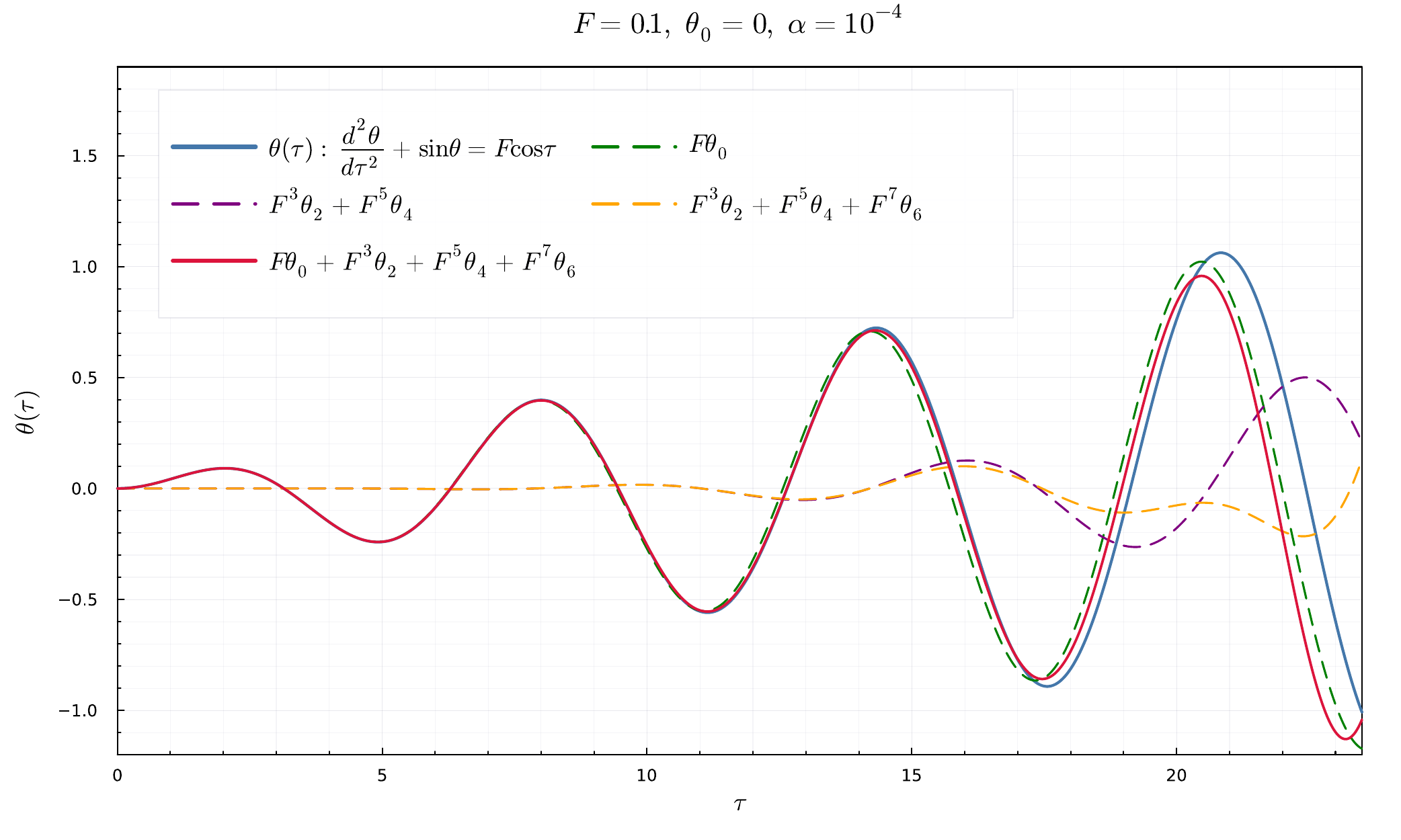}
	\caption{Comparison of the obtained solutions at various orders with the actual solution}
   \label{figf3}
\end{figure*}
As with $\theta_{1}$, this amplitude clearly goes to 0 as well. Therefore, we go to the next order,
\begin{equation}\nonumber
    \Ddot{\theta}_{4}(\tau) \ + \ \theta_{4}(\tau) \ = \ \frac{1}{2}\theta_{0}^{2}\theta_{2} \ = \ \frac{-\tau^{4}}{4096}\Big(\alpha + 
    \frac{\tau}{2}\Big)^{2}\cos\tau 
\end{equation}
\begin{equation}\ = \ \Tilde{G}(\tau) 
\end{equation}
We approximate and keep the largest term, $\Tilde{G}(\tau) \sim - a_{1}\tau^{6}\cos\tau$, and as above, we have ($a_{1}$ is the constant numerical prefactor in the 
expression),
\begin{equation}\nonumber \theta_{4}(\tau) \ = \ \end{equation}
\begin{equation}\nonumber \sin{\tau}\int_{0}^{\tau}\Tilde{G}(\xi)\cos{\xi} \,d\xi -  \cos{\tau}\int_{0}^{\tau}\Tilde{G}(\xi)\sin{\xi} \,d\xi\end{equation}
\begin{equation} \ = \ J_{1}\sin\tau  - J_{2}\cos\tau \end{equation}
The integrals yield (keeping only the largest term), 
\begin{equation}J_{1} \ \sim \ -\frac{8a_{1}}{112}\tau^{7}; \ J_{2} \ \sim \ \frac{a_{1}}{4}\tau^{6}\cos2\tau \end{equation}
Or, 
\begin{equation}\label{four}
    \theta_{4}(\tau) \ \sim \ - \frac{a_{1}}{14}\tau^{7}\sin\tau 
\end{equation}
The sign of the amplitude is still negative and exponentiation of time larger than earlier and the increase in amplitude, therefore, is much faster. The next order 
again provides no contribution like $\theta_{1}$, but the one after that features a significant change. The equation in $\mathcal{O}(F^{7})$ is, 
\begin{equation}\nonumber
    \Ddot{\theta}_{6}(\tau) \ + \ \theta_{6}(\tau) \ = \ \frac{1}{2}\theta_{0}\theta_{2}^{2} \ + \ \frac{1}{2}\theta_{0}^{2}\theta_{4} \ = 
\end{equation}
\begin{equation} \label{req}
    \chi_{1}(\tau) \ + \ \chi_{2}(\tau) \ \sim \ a_{2}\tau^{9}\sin\tau \ = \ \Bar{G}(\tau)
\end{equation}
As before, this equation leads us to, 
\begin{equation}\nonumber \theta_{6}(\tau) \ 
= \ K_{1}\sin\tau  - K_{2}\cos\tau 
\end{equation}
The largest term from the above integrals is due to $K_{2} \ \sim \ \frac{1}{20}a_{2}\tau^{10}$. This leads to,
\begin{equation}\label{six}
    \theta_{6}(\tau) \ \sim \ -\frac{a_{2}}{20}\tau^{10}\cos\tau
\end{equation}
The interesting part about this solution is its sign - the amplitude can grow with time \textit{if} the numerical prefactor is negative - which it is. If we go back 
to \eqref{req}, $a_{2}$ is like the difference in the 2 different forcing terms and its sign depends on which of the forcing terms is larger. \\
Fig. (\ref{figf3}) depicts the solutions obtained using the perturbative approach (the pink curve) and compares them with the actual solution (the blue curve) 
obtained through numerical integration - there is a remarkable match between the two for an initial amount of time (the approximation gets better for smaller times 
because the linear order dominates). It is clear that successively adding higher order terms causes modifications to the amplitude of the total solution. This is 
especially clear around the $\tau \approx 19$ and $\tau \approx 23$ points, where one can see that adding the $\mathcal{O}(F^{7})$ term to the lower order solution 
(the violet curve) causes, in one case, a maxima to become a minima and, in the other, suppresses the magnitude of the amplitude (the yellow curve). Further, it is 
also clear that for the higher order corrections considered here, the perturbative solution is a better fit to the actual than the linear approximation for until 
$\tau \approx 20$ - specifically, in the interval of $\tau \in [11, 20]$, the pink curve is closer to the actual solution than the linear approximation. \\
Further, in Fig. (\ref{figu1}), we see that the trigonometric terms imposing the phase difference are important for the solution to work. The plot depicts the same 
$n^{th}$ degree polynomials in $\tau$, $p_{n}$, that form the coefficients of the trigonometric terms of the solution in Fig. (\ref{figf3}). Their impact is most 
easily seen by comparing the $\tau = 22$ mark in Figs. (\ref{figf3}) and (\ref{figu1}) - in the latter case,
the function crosses $\theta = 1$ and grows further, while in the former case, both it takes on $\theta \approx 0$. This shows that successive higher order terms can 
cause suppression of the amplitude of the total system and may be able to damp the diverging behaviour of the individual solutions themselves even in the absence of 
external damping mechanisms. 

\subsubsection{Bounds}\label{sect3.1}
As a further check, one can easily consider a simple bound of the solution thus obtained. In general, the $n^{th}$ order solutions are zero if $n$ is odd and 
non zero if even. Therefore, the $n^{th}$ order amplitude can be written as (for $n = 0, 2, 4...$),
\begin{equation}
    |A_{n}| \ = \ \tau^{n/2}(\tau F)^{n+1}\Big|\sin{(\tau + \frac{n \pi}{4})}\Big|
\end{equation}
The total amplitude is bounded from above by,
\begin{equation}
    S(\tau) = \sum_{n=0}^{\infty} |A_{2n}| = \sum_{n = 0}^{\infty} \tau^{n}(\tau F)^{2n+1}\Big|\sin{(\tau + \frac{n \pi}{2})}\Big|
\end{equation}
This is a geometric series, and so, 
\begin{equation}
    S(\tau) = \tau F\Big[|\sin{\tau}|\sum_{m = 0}^{\infty}a^{2m} + |\cos{\tau}|\sum_{m = 0}^{\infty}a^{2m+1}\Big]
\end{equation}
Here, $a = \tau^{3}F^{2} < 1$ for validity. Thus,
\begin{figure}[H]
    \centering
    \includegraphics[width = \columnwidth]{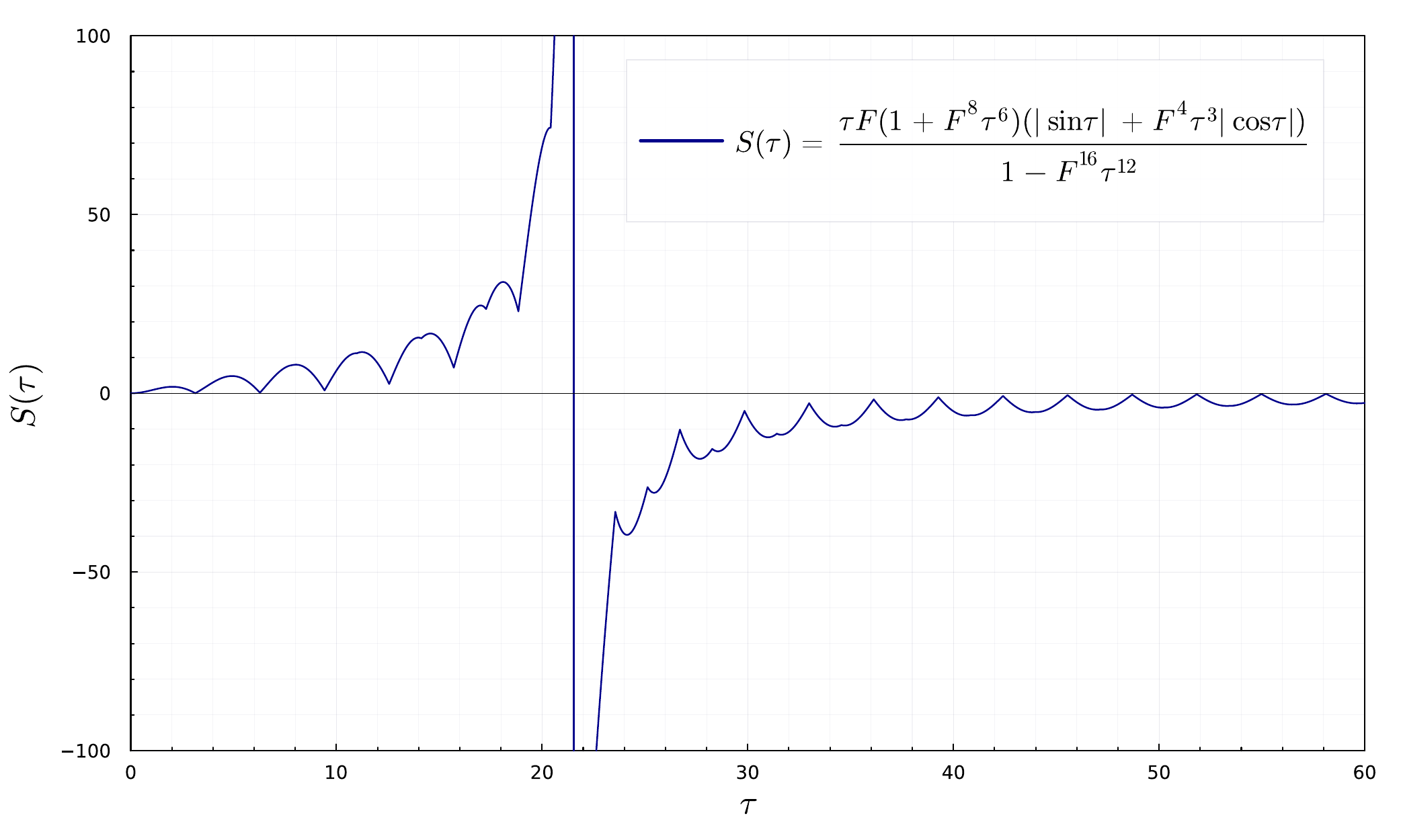}
    \caption{The upper bound on the answer}
    \label{figbound}
  \end{figure}  
\begin{equation}\label{bound}
    S(\tau) \ = \ \tau F \frac{|\sin{\tau}| + \tau^{3}F^{2}|\cos{\tau}|}{1 - (\tau^{3}F^{2})^{2}}
\end{equation}
Eq. \eqref{bound} is plotted with respect to $\tau$ in Fig. \ref{figbound} - As one can see, the expression is quite small and rises slowly till the divergence 
point. The calculations above are only valid up to the point of divergence - however, from Eq. \eqref{bound} it is clear that $F$ determines the divergence point 
and the smaller the driving force, the later is the divergence. This means that the validity of our perturbative approach increases (and the uppper bound remains 
small and finite) for longer times as the driving force becomes smaller. \\
For instance, for $F \sim 0.001$, this solution remains valid till around $\tau \sim 100$ - 
quite long. It should be noted that the above calculation provides a strict upper bound (since there is no general expression for the phase differences in every 
order) - therefore, the divergence point bears no real meaning for the actual oscillator. However, the upper bound being small and finite before the divergence 
point does imply that the real answer must be small and finite as well. 

\section{Traditional perturbation theory: successes and failures}\label{sect2}
\begin{figure}[H]
	\includegraphics[width = \columnwidth]{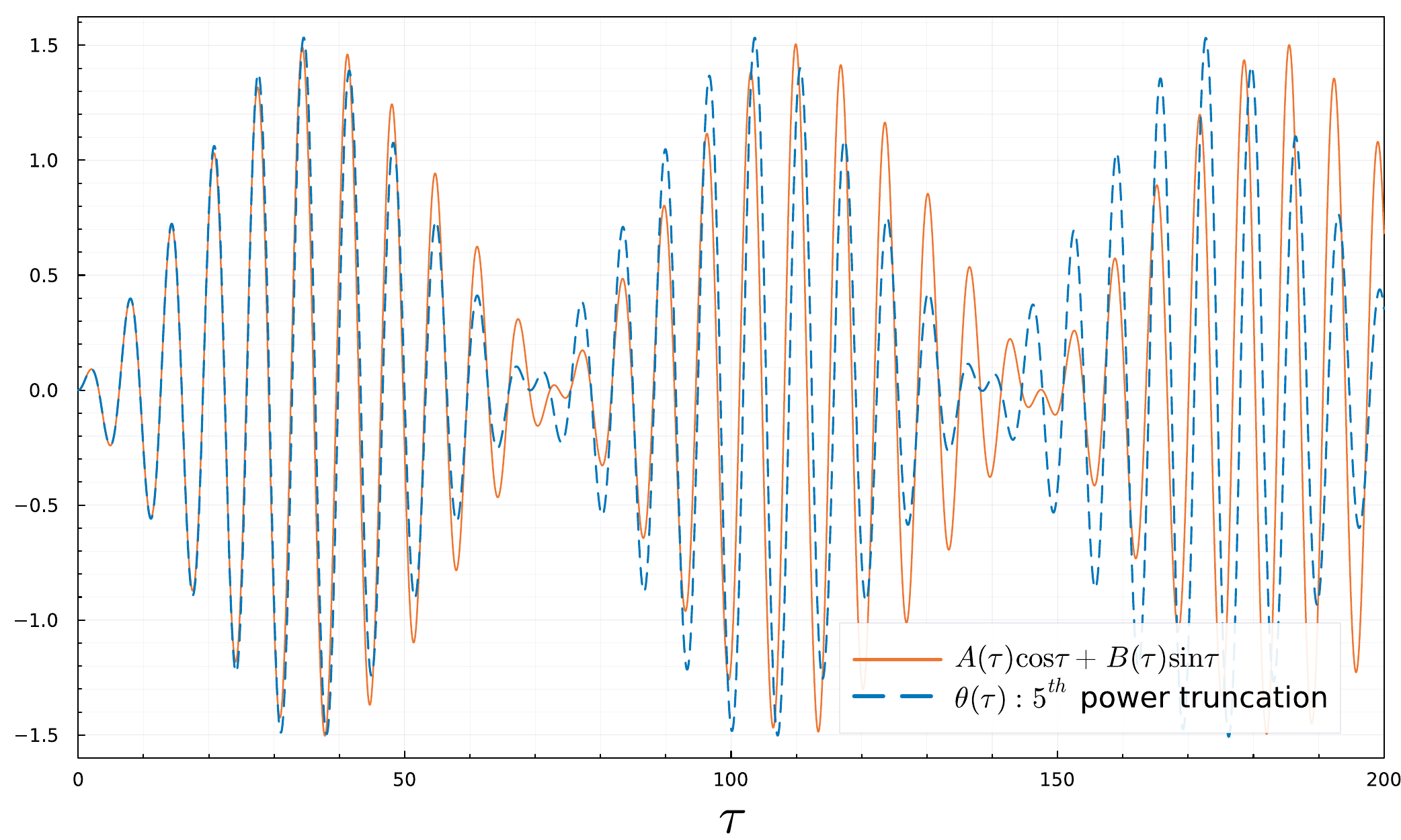}
	\caption{Comparison of the true solution ($\theta^{5}$ truncation in eqn. \eqref{eq3.1}) with the solution described by eqns. \eqref{KB1} and \eqref{KB2}}
  \label{figKB}
\end{figure}

In the previous section, we have introduced a different form of perturbation theory which is capable of giving very accurate answers at short times. Now, in the 
present section, we fall back on traditional techniques which are geared to explorinig the long time limit of the dynamics. We believe that combined with the short
time picture from the previous section, we have a very reasonable picture of the dynamics over a long period of time. \\
To start with, it is interesting to note that a naive perturbative expansion does not work well for the problem at hand. Specifically, if we expand as, 
\begin{equation}
    \theta = \theta_{0} + H\theta_{1} + H^{2}\theta_{2} + ...
\end{equation}
Then, 
\begin{equation}\nonumber
    \sin \theta = \sum_{n = 0}^{\infty} \frac{(-1)^{n}}{(2n+1)!}\theta^{2n+1} 
\end{equation}  
\begin{equation}\nonumber
 = \sin \theta_{0} + (H\theta_{1} + H^{2}\theta_{2} + ...)\cos \theta_{0} 
\end{equation}
\begin{equation}
    +\ (H^{2}\theta_{1}^{2} + H^{4}\theta_{2}^{2} + ...)\sin \theta_{0} + ...
\end{equation}
This implies that provided $\theta$ is expanded as above, we can never get rid of the non-linearities using the usual perturbation techniques. We work around this 
using a very specific expansion in section \ref{sect3}. \\
However, even if a naive perturbative expansion fails, there are other approaches that can be used to investigate the problem in more detail. We begin by expanding 
the sine in a Taylor series and rescaling the problem to make things dimensionless - we denote the rescaled time by $\tau = \omega t$ and redefine $H = \omega^{2}F$ 
to obtain, 
\begin{equation} \label{eq3.1}
    \frac{d^{2}\theta}{d\tau^{2}} + \big(\theta +\frac{1}{6} \theta^{3} +\frac{1}{120} \theta^{5}+\dots\big) = F\cos\tau
\end{equation}
Our procedure will be to do an effective linearization and write the LHS of \eqref{eq3.1} as $\Ddot{\theta}+\Omega^{2}\theta$ (with the dotted derivative now denoting
a derivative with respect to $\tau$) where $\Omega^{2}$ will be found as an expansion in the amplitude of motion, $A$. We will employ both the Linstedt-Poincaré and
Krylov-Bogoliubov techniques and show that the same answer is provided by very different approaches. This is indicative of the robustness of the perturbative 
approach. \\
We begin by writing \eqref{eq3.1} as, 
\begin{equation}
    \Ddot{\theta} + \theta + \lambda\theta^{3} + \mu\theta^{5} = F\cos\tau 
\end{equation}
We shall use the coefficients $\lambda$ and $\mu$ as perturbation parameters. We start with the Linstedt-Poincaré technique - as stated earlier, we wish to express
the LHS of \eqref{eq3.1} in a specific form where the effective frequency, $\Omega$, will capture the effects of the cubic and quintic terms. The determination of 
$\Omega$ does not depend on the specific form of the driving term (i.e., the RHS of \eqref{eq3.1}) and our goal will be to consider the anharmonic oscillator with
a potential of the followed form,
\begin{equation}\label{eq3.2}
V(\theta) = \frac{\theta^{2}}{2} + \frac{\lambda}{4}\theta^{4} + \frac{\mu}{6}\theta^{6}
\end{equation}
We will aim to derive an expression for the period of oscillations in terms of the amplitude, $A$. Therefore, we look at the unforced dynamics, $\Ddot{\theta} + 
dV/d\theta = 0$, by expanding $\theta(\tau)$ as, 
\begin{equation}\label{eq3.3}
    \theta(\tau) = \theta_{0} + (\lambda \theta_{1}+\lambda^{2}\theta_{2}+\dots) + (\mu\theta'_{1}+\mu^{2}\theta'_{2}+\dots)
\end{equation}
Then, the unforced equation takes the form, 
\begin{equation}\label{eq3.4}
    \Ddot{\theta} + \frac{dV}{d\theta} = \Ddot{\theta} + \omega^{2}\theta + (1-\omega^{2})\theta + \lambda\theta^{3} + \mu\theta^{5} = 0
\end{equation}
Here, $\omega$ denotes the true frequency of the oscillation in the potential, $V(\theta)$. We expand it as, 
\begin{equation}\label{eq3.5}
    \omega^{2} = 1 + (\lambda \omega_{1}+\dots) + (\mu\omega'_{1}+\dots)
\end{equation}
With the expansions in \eqref{eq3.3} and \eqref{eq3.5}, the oscillator in \eqref{eq3.4} gives in the various perturbative orders, 
\begin{equation}
    \mathcal{O}(0):\ \Ddot{\theta}_{0} + \omega^{2}\theta_{0} = 0
\end{equation}
\begin{equation}
    \mathcal{O}(\lambda):\ \Ddot{\theta}_{1} + \omega^{2}\theta_{1} = \omega_{1}^{2}\theta_{0}-\theta_{0}^{3}
\end{equation}
\begin{equation}
    \mathcal{O}(\mu):\ \Ddot{\theta}'_{1} + \omega^{2}\theta'_{1} = \omega_{1}'^{2}\theta_{0}-\theta_{0}^{5}
\end{equation}
\begin{equation}
    \mathcal{O}(\lambda^{2}):\ \Ddot{\theta}_{2} + \omega^{2}\theta_{2} = \omega_{2}^{2}\theta_{0}-3\theta_{0}^{2}\theta_{1}
\end{equation}
Thus, we have $\theta_{0} = A\cos\omega\tau$ (assuming that $A$ is the initial displacement and the initial velocity is $0$). Substituting this in the other equations,
we find, 
\begin{equation}
    \Ddot{\theta}_{1} + \omega^{2}\theta_{1} = \omega_{1}^{2}A\cos\omega\tau-\frac{A^{3}}{4}(3\cos\omega\tau+\cos3\omega\tau)
\end{equation}
To remove the spurious resonance on the RHS due to the $\cos\omega\tau$ terms, we find that $\omega_{1}^{2}=3A^{2}/4$ and hence, the $\mathcal{O}(\lambda)$ correction
to the response is given as, 
\begin{equation}
    \omega^{2} = 1 + \frac{3\lambda}{4}A^{2}
\end{equation}
We can similarly deduce the $\mathcal{O}(\mu)$ and $\mathcal{O}(\lambda^{2})$ corrections to the response - the resonance inducing terms arise from the 
$\theta_{0}^{5}$ and $3\theta_{0}^{2}\theta_{1}$ terms, 
\begin{equation}
    \omega^{2} = 1 + \frac{3\lambda}{4}A^{2} + \Bigg(\frac{5\mu}{8} + \frac{3\lambda^{2}}{128}\Bigg)A^{4} + \dots
\end{equation}
We now use $\lambda = -1/6$ and $\mu = 1/120$. Then the effective linearized equation of motion corresponding to \eqref{eq3.1} is given by, 
\begin{equation}
    \Ddot{\theta} + \Bigg(1-\frac{A^{2}}{8}+\frac{3A^{4}}{512}\Bigg)\theta = F\cos\tau
\end{equation}
The solution to the above is of the form, 
\begin{equation}
    \theta(\tau) = -\frac{F}{\frac{A^{2}}{8}-\frac{3A^{4}}{512}}\cos\tau
\end{equation}
Self consistency with the original ansatz demands that, 
\begin{equation}\label{compare}
    |A| = \frac{8F}{A^{2}(1-\frac{3A^{2}}{64})}
\end{equation}
For small $F$, this implies,
\begin{equation}\label{prop}
    |A| \propto F^{1/3},\ F << 1
\end{equation}
\begin{figure}[H]
    \includegraphics[width = \columnwidth]{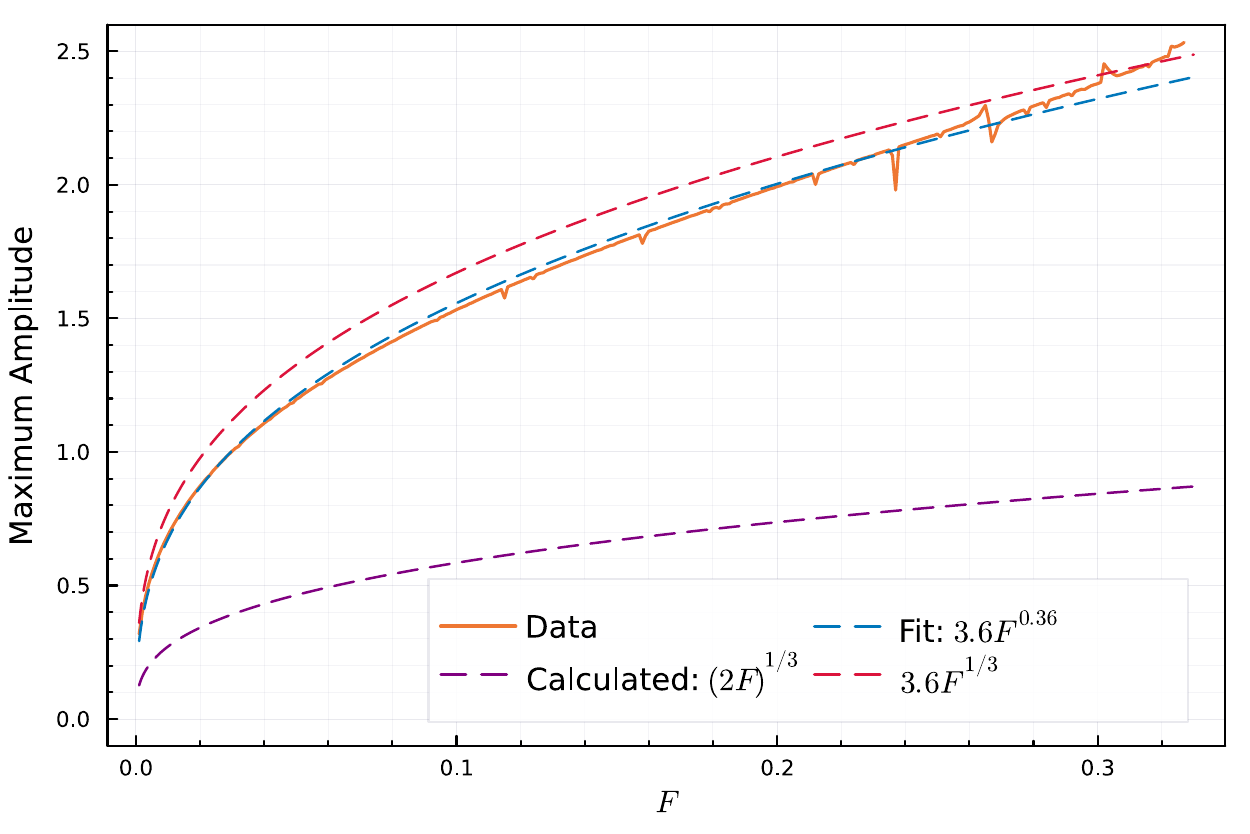}
    \caption{The drive-response relation}
    \label{figdrive}
\end{figure}
The drastic nature of the process (where the amplitude dependence of the frequency was calculated neglecting the external forcing term and reinstating it after
the calculation) makes our result trustworthy for gross features like the existence of a solution as in \eqref{prop} but cannot ensure the stability of such a 
solution. This also matches our numerical result in Fig. \ref{figdrive}. While the prefactor in our rough estimation of \eqref{prop} is uncertain, 
the exponential factor of $1/3 \approx 0.33$ is not very far off from the actual behaviour of the simulated system.\\ 
We now turn to the Krylov-Bogoliubov scheme to obtain a somewhat more detailed picture of the dynamics with the driving term playing an essential part in the
calculation this time. We begin by choosing the ansatz,
\begin{equation}\label{KB0}
    \theta(\tau) = A(\tau)\cos\tau + B(\tau)\sin\tau
\end{equation}
Here, $\tau = \omega t$ as before and $A,\ B$ are slowly varying functions of $\tau$, implying that $\Ddot{A},\ \Ddot{B}$ are much smaller than $A,\ B$ and can be 
subsequently ignored. A comparison of the numerical solution based on these assumptions and the actual solution is presented in Fig. \ref{figKB} - as we can see, the
there is a good overlap of the solutions for at least $\tau \sim 200$. Now, substituting eqn. \eqref{KB0} in \eqref{eq3.1}, we have, 
\begin{equation}\label{KB1}
\Dot{A} = -\frac{B}{16}\Big(A^{2}+B^{2}\Big) + \frac{B}{384}\Big(A^{2}+B^{2}\Big)^{2}
\end{equation}
\begin{equation}\label{KB2}
\Dot{B} = \frac{F}{2} +\frac{A}{16}\Big(A^{2}+B^{2}\Big) - \frac{A}{384}\Big(A^{2}+B^{2}\Big)^{2}
\end{equation}
Changing variables to $r^{2} = A^{2}+B^{2}$ and $\tan\phi = B/A$ takes us to a more convenient form, 
\begin{equation}\label{rdot}
    \Dot{r} = \frac{F}{2}\sin\phi
\end{equation}
\begin{equation}\label{phidot}
    \Dot{\phi} = \frac{1}{2r}\Big[F\cos\phi +\frac{r^{3}}{8}\Big(1 - \frac{r^{2}}{24}\Big)\Big]
\end{equation}
To get the essence of the above dynamics, we adopt the dynamical systems approach and look for fixed points and their stabilities for the above system. The fixed 
points are, for one variable, $\phi^{*} = 0,\ \pi$. This gives us two equations, 
\begin{equation}\label{fpt1}
    F\pm\frac{r^{*3}}{8}\Big(1 - \frac{r^{*2}}{24}\Big) = 0
\end{equation}
For $F>0$, only the second case ($\phi^{*} = \pi$) is relevant. It is interesting to note that eqns. \eqref{compare} and \eqref{fpt1} are nearly identical in 
structure. The numerical investigation in section \ref{sect4} shows that the dynamics becomes more involved after $F \sim 0.33$, so our range of interest will be 
for $F$ significantly less than unity. Now, for $F << 1$, we see that $r^{*}$ is very close to $8F$ - therefore, we have an amplitude close to $2F^{1/3}$. We now 
proceed further and investigate whether $r^{*} \approx 2F^{1/3}$ is a stable fixed point or not. Accordingly, we expand eqns. \eqref{rdot} and \eqref{phidot} around 
$\phi^{*}=\pi$ and the corresponding $r^{*}$ as obtained from \eqref{fpt1} - essentially, this means linearizing eqns. \eqref{rdot} and \eqref{phidot} by using 
$\phi = \pi+\delta\phi$ and $r = r^{*}+\delta r$. This gives us, 
\begin{equation}
    \delta \dot{r} = -\frac{F}{2}\delta\phi
\end{equation}
\begin{equation}
    \delta\dot{\phi} = \frac{3r^{*}}{16}\Bigg(1-\frac{5r^{*^{2}}}{72}\Bigg)\delta r
\end{equation}
\begin{figure}[H]
    \includegraphics[width = \columnwidth]{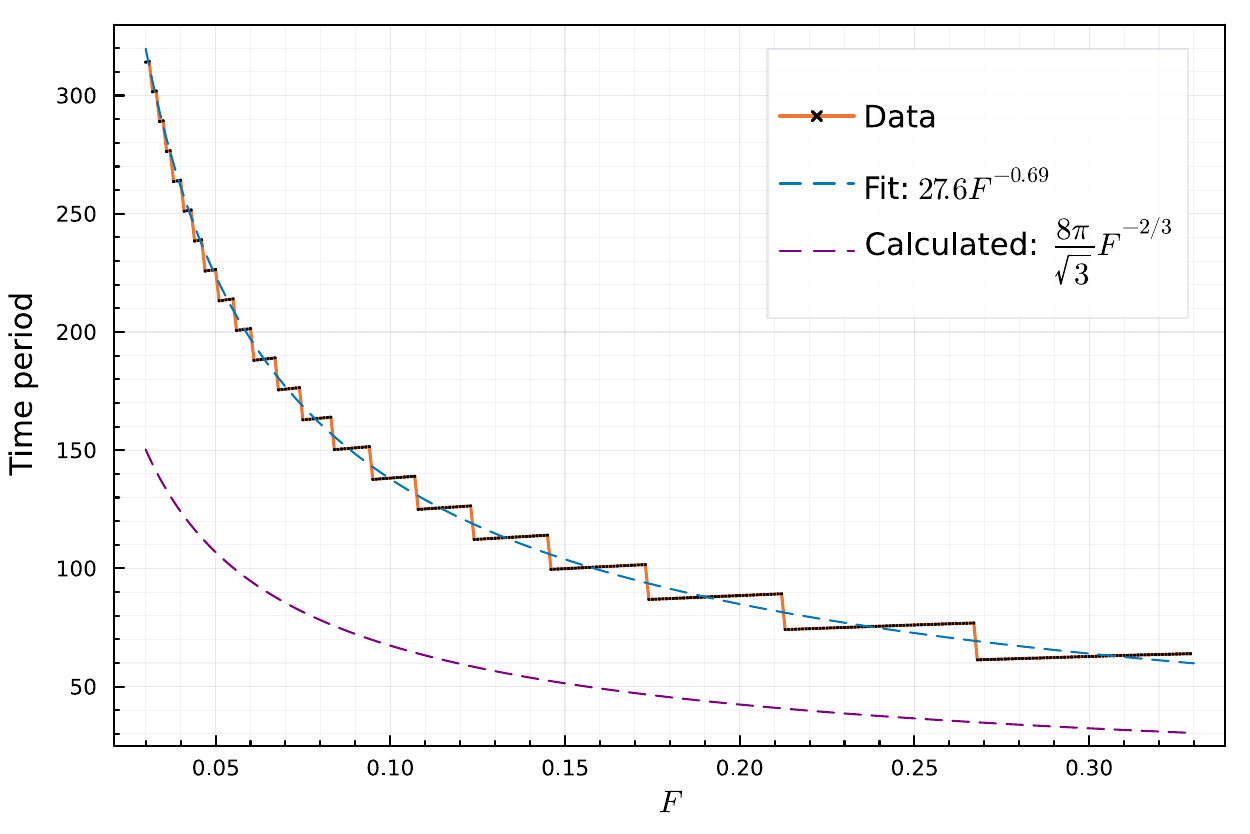}
    \caption{The $F$ dependence of the slower modulation period}
    \label{figtime}
\end{figure}
This implies, 
\begin{equation}
    \delta \Ddot{r} + \frac{3F}{32}r^{*}\Bigg(1-\frac{5}{72}r^{*^{2}}\Bigg)\delta r = 0 
\end{equation}
Now, since $r^{*} \approx 2F^{1/3}$, this reduces to, 
\begin{equation}
    \delta \Ddot{r} + \frac{3F^{4/3}}{16}\Bigg(1-\frac{5}{18}F^{2/3}\Bigg)\delta r = 0 
\end{equation}

\begin{figure*}
	\includegraphics[width = \textwidth]{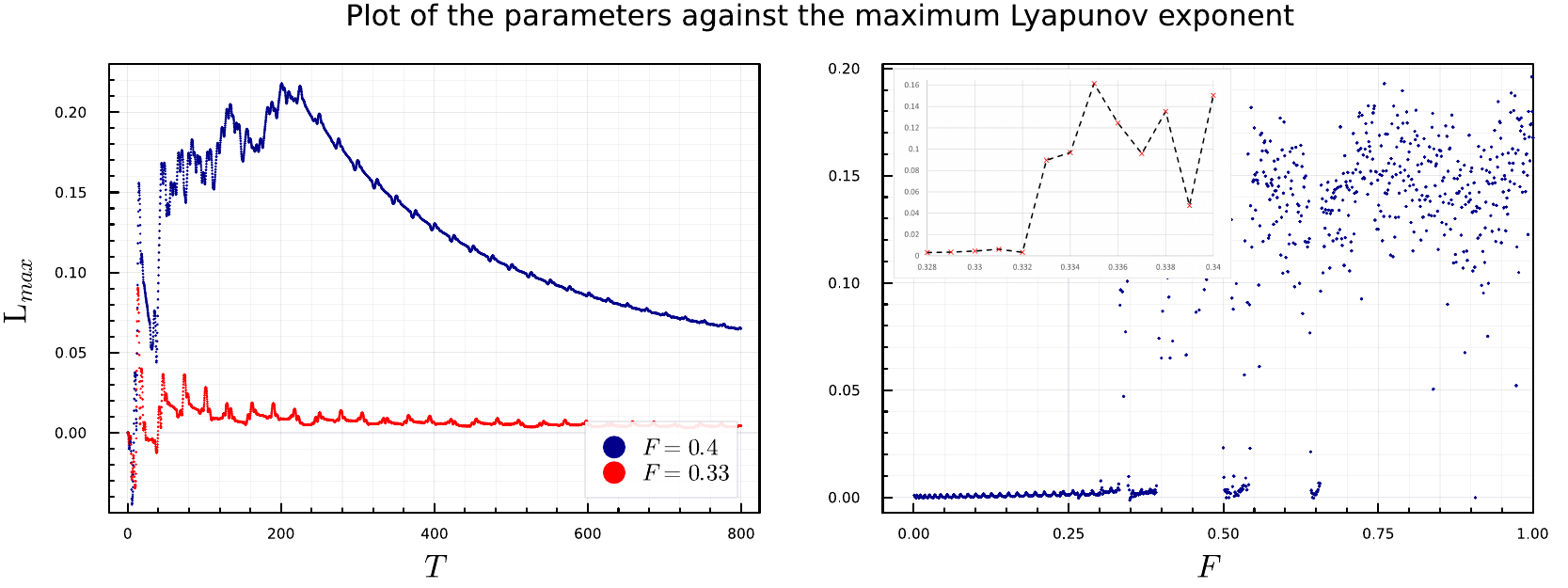}
	\caption{Variation of the maximum Lyapunov exponent with parameters}
   \label{figlya}
\end{figure*}
For $F<<1$, this implies, 
\begin{equation}
    \delta \Ddot{r} + \frac{3F^{4/3}}{16}\delta r = 0
\end{equation}
This reveals that the fixed point under consideration here is a centre and the solution is approximately given by, 
\begin{equation}
    r(\tau) = 2F^{1/3} + C\cos\Bigg[\frac{\sqrt{3}F^{2/3}}{4}\sqrt{1-\frac{5}{18}F^{2/3}} \tau\Bigg]
\end{equation}
Here, $C$ is the integration constant. For small $F$, this becomes, 
\begin{equation}
    r(\tau) = 2F^{1/3} + C\cos\Bigg[\frac{\sqrt{3}F^{2/3}}{4}\tau\Bigg]
\end{equation}
Therefore, the amplitude of our driven undamped oscillator is of the form $C\cos\frac{\sqrt{3}F^{2/3}}{4}\tau$ where $C$ is an undetermined 
constant at this level of approximation. This is manifested in our numerical simulations which depict two clear timescales - in Fig. \ref{figtime}, we plot 
the dependence of the slower modulation period against $F$ and the $F^{-2/3}$ dependence of the period can be clearly seen.\\
We reiterate that the two results obtained here - one, that the resonant amplitude behaves as $F^{1/3}$ and, two, that the modulation frequency of the envelope is 
$F$-dependent (proportional to $F^{2/3}$) - are independent of the order of calculation and provide an interesting insight into the nature of the system. These 
calculations are only approximate and we emphasize that the scaling exponents are more important than their coefficients - including higher order terms can only 
modify the coefficients. Such exponents derived through self consistency arguments often turn out to be exact \cite{halp}, while the amplitude requires a 
perturbative expansion that only slowly converges.

\section{Onset of chaos}\label{sect4}
\begin{figure}[H]
    \includegraphics[width = \columnwidth]{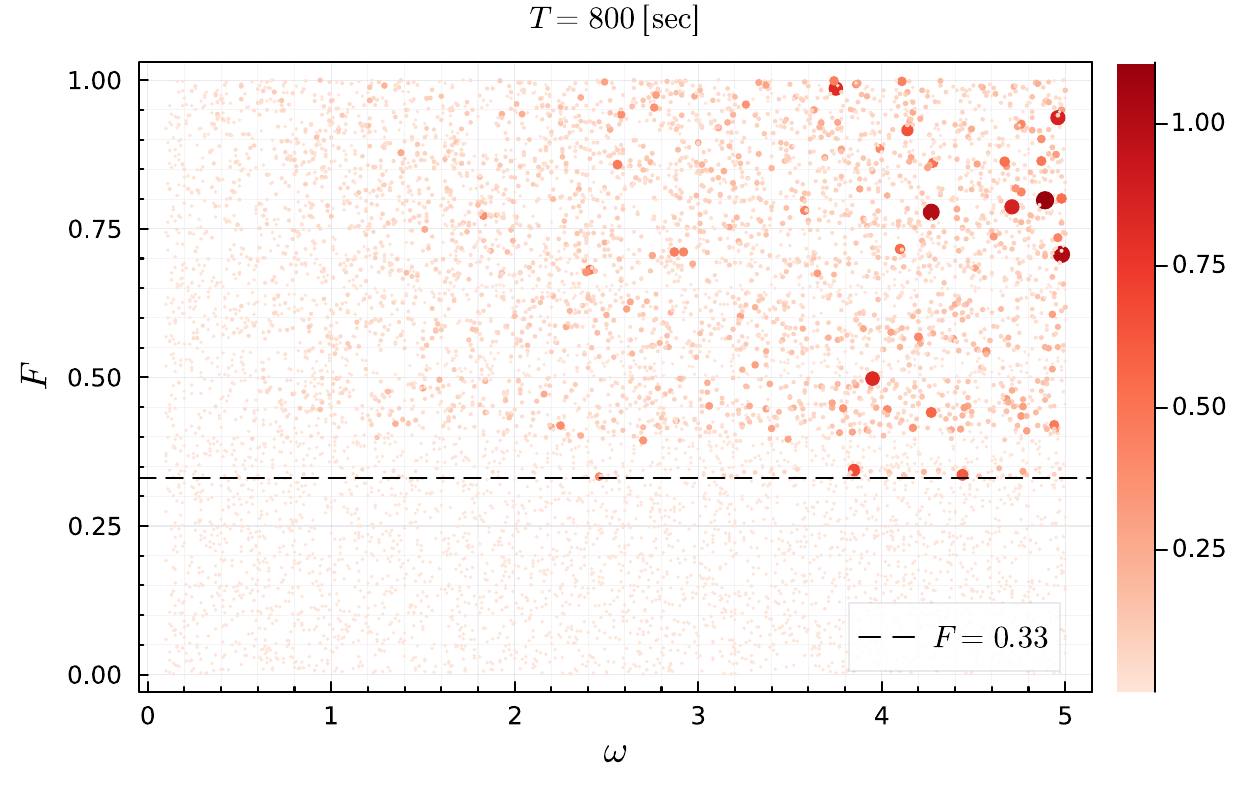}
    \caption{Long time behaviour of the response in the parameter space}
    \label{figidk}
\end{figure}

While a seemingly simple system, this undamped forced oscillator presents an interesting behaviour if the forcing amplitude is varied - in section \ref{sect2}, 
we showed how the maximum amplitude and the slower modulation period both scale with the drive. Till now, our analysis only 
considered a small forcing amplitude, but what if that were not the case? A numerical study into the system shows us that the dynamics becomes more varied as one 
increases the forcing strength - $F \sim 0.33$ seems to be the point beyond which this becomes more apparent.\\
The essence of this is captured in the plots depicted in Fig. \ref{figlya}, showing how the maximum Lyapunov exponent varies with the dimensionless drive, $F$, and 
the total time of evolution, $T$ \cite{Bennetin}\cite{balcerzak}. The plot displays results sampled from $8000$ trials with $\omega$ fixed at $1\ [rad/s]$ while 
$F$ was generated randomly from $0.001 - 1$ with a precision of $0.001$. For all the trials, the system is evolved for a total time of $800\ [s]$. Fig. \ref{figidk},
on the other hand, shows how the long time amplitude of the oscillation varies in the parameter space. It also consists of points sampled from $8000$ trials with
both $\omega$ and $F$ generated randomly - the former in the range $0.1 - 5\ [rad/s]$ with a precision of $0.01\ [rad/s]$ and the latter in the range mentioned
earlier. The colour and size of the points are proportional to the magnitude of the response (normalised by the total number of trials, $8000$) - both a 
darker colour and bigger size indicate a larger response (the total time of evolution in this case was $800 \ [s]$). 
\begin{figure*}
	\includegraphics[width = \textwidth, height = 55mm]{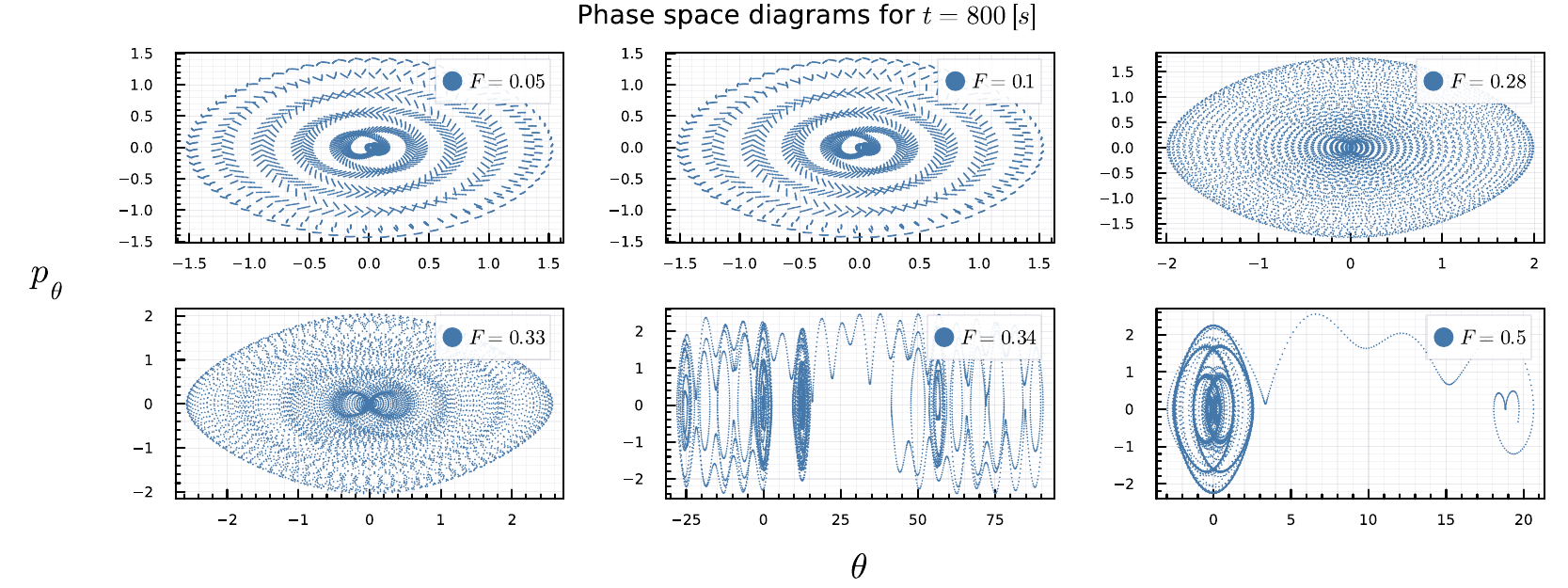} 
        \includegraphics[width = 0.8\textwidth, height = 55mm]{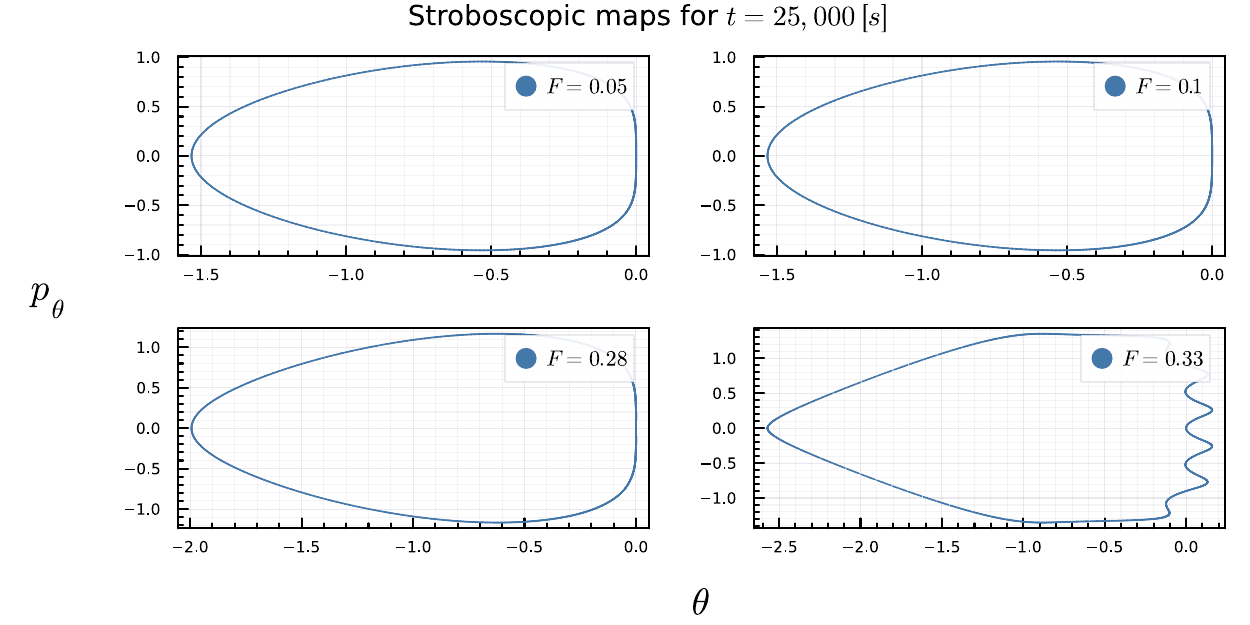}
        \caption{Phase space maps for different choices of $F$}
        \label{figPM}
\end{figure*}
The variation against the \textit{dimensionless} $F$ makes it clear that the exponent is exceedingly small for very small values of $F$ from $0.001$ to just over 
$0.3$ - this is followed by a steep and sudden increase, as is evident from the inset, after which the exponent again randomly fluctuates around 
$\langle L_{max}\rangle \approx 0.12$. The Lyapunov exponent, of course, is not very big for the selected range - but the relative difference in the region before 
and after $F \sim 0.33$ is certainly interesting and suggests a transitional behaviour. Rather than a slow build up, the plot shows more of a step function kind of 
behaviour indicative of a first order transition dependent on the drive. The plot on the left depicts two curves for $F = 0.33,\ 0.4$ - consistent with the 
one on the right, both show a saturating behaviour as $T$ becomes larger but with a very noticeable difference in the Lyapunov values. Similar investigations on 
the scaling of the Lyapunov exponents with time periods have been carried out earlier in the context of other systems(see \cite{Stoop}). Similarly, in Fig. 
\ref{figidk}, all of the larger responses are for $F> 0.33$ (the dotted black curve). \\
These conclusions are more obvious from an analysis of the phase space diagrams, specifically the Poincaré maps of the system for these parameters, depicted in 
Fig. \ref{figPM}. The first set of diagrams (the upper block of six plots) depict depicts the phase space diagrams for the various choices of the dimensionless $F$ 
values for a total time of evolution of $800\ [s]$. The lower set depicts Poincaré maps for some of those choices of $F$ when the system is evolved for a total time 
period of $25,000\ [s]$ with the sampling rate being the usual $2\pi\ [s]$.  \\
\begin{figure*}
    \begin{minipage}{\columnwidth}
        \includegraphics[width=0.99\columnwidth]{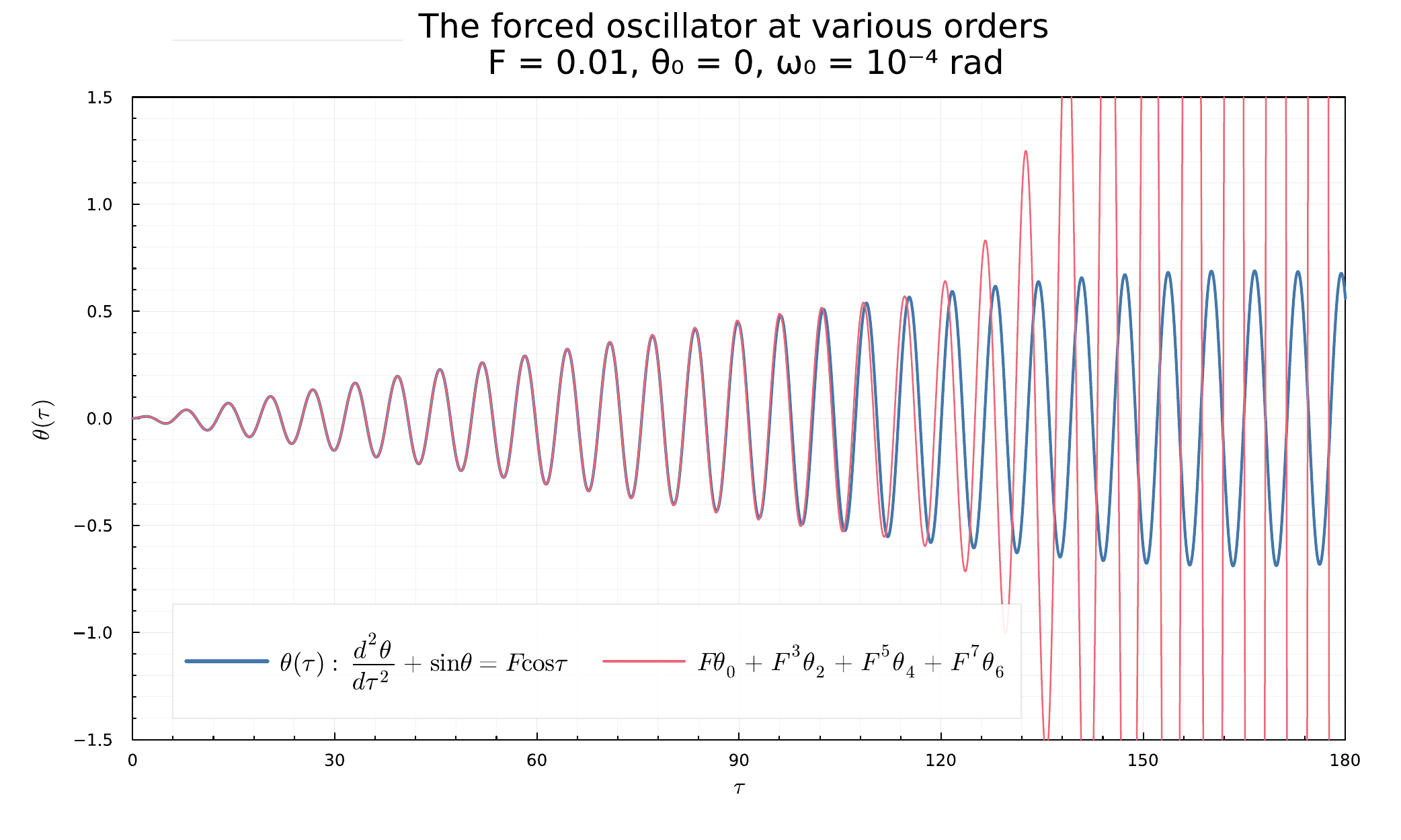}
    \end{minipage}
    \begin{minipage}{\columnwidth}
        \includegraphics[width=0.99\columnwidth]{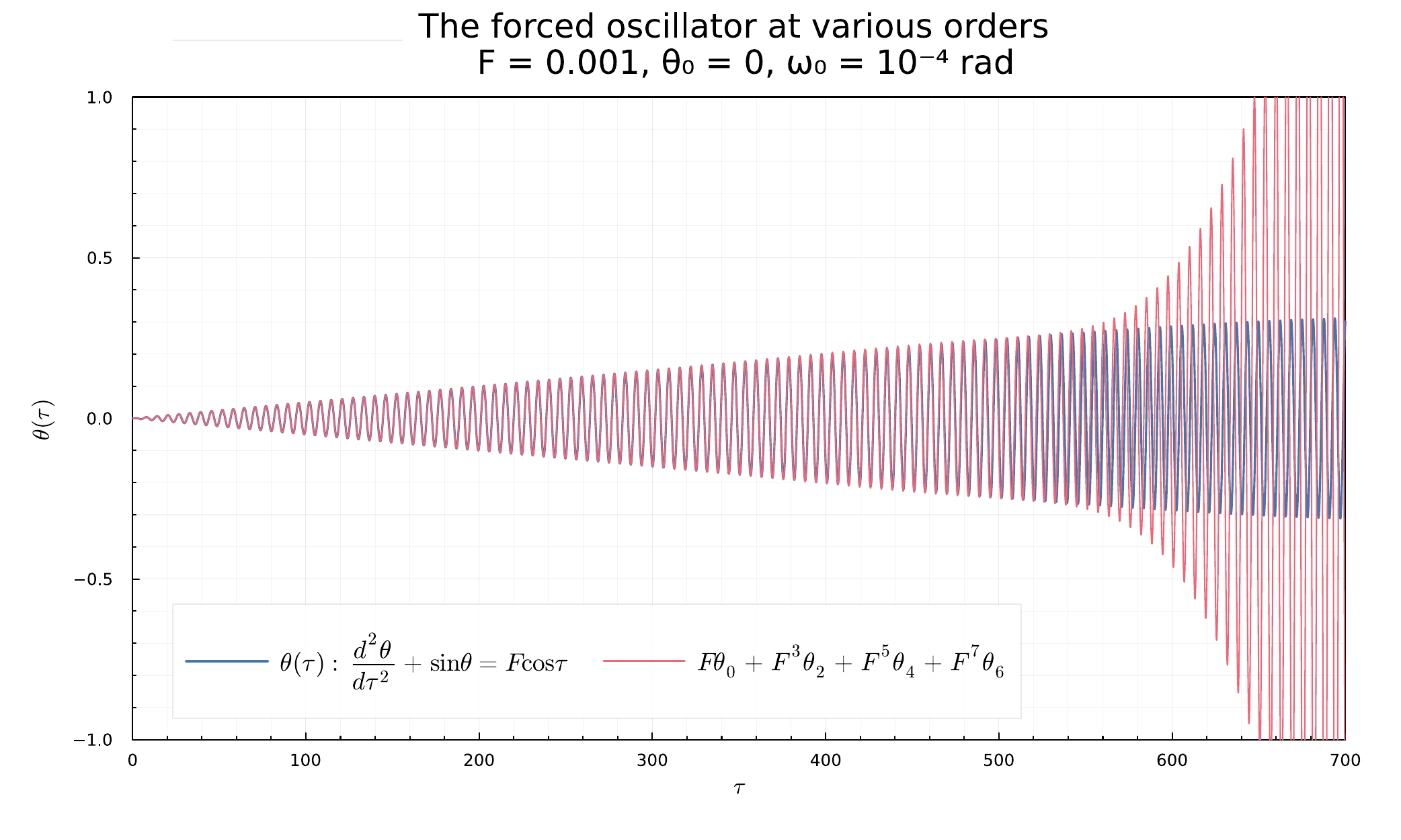}
    \end{minipage}
     \caption{Comparison of the exact and approximate solutions of the forced oscillator for different $F$}
         \label{figdiff}
\end{figure*}
From the Poincaré maps, it is clear that the smaller values of $F$ correspond to closed orbits. An increase in the value of $F$ seems to affect this periodicity 
which is broken when $F$ becomes greater than an approximate value of $0.33$. This is exactly the kind of transitional behaviour we expected from Fig. \ref{figlya}. 
It is clear from the Poincaré maps, that as $F$ increases the orbit becomes more undulated, signifying a larger time period - this is a large $F$ effect, however, 
and not privy to the discussion in section \ref{sect2}. \\
The numerical simulation was observed to be algorithm dependent for large times for values of $F$ beyond 
$0.33$ (believed to be an indication of the numerical chaotic behaviour) - we have checked several values both below $0.33$ and above, and we haven't yet observed 
such an algorithm dependence for large time scales for the ones less than the critical point. This is the reason we have not depicted Poincaré maps for 
$F = 0.34,\ 0.5$ since we were unable to verify the accuracy of the methods. The phase space maps were, however, adjusted properly with lower error tolerance 
values than usual. 

\section{Discussion}\label{sect5}
We have tried to investigate finite resonant solutions perturbatively in an undamped forced oscillator by assuming a small forcing amplitude. Following the 
perturbative expansion introduced in section \ref{sect3}, we have derived a series of equations with various forcing terms for different orders. We have looked into 
the first few orders in detail, noting that there is a possibility for different orders to differ in phases. We believe that this can result in a final bounded 
solution despite the fact that individual solutions at each order may diverge. This idea is strikingly similar to destructive interference where differing phases 
between different waves can cause a suppression of the total amplitude.\\
Interference is a ubiquitous idea in physics - the basic idea lies in the different phases of superposing waves. Two sine waves half a wavelength out of phase 
cancel each other out. Such a sine wave is represented by $\sin{(\theta + \pi)} = -\sin{(\theta)}$; therefore, the sign difference signifies the differing phases 
in the two waves. Something similar is observed here with differing signs in alternative orders - the out of phase solutions in different orders lead to a 
cancellation (or, a suppression) of the individual diverging amplitudes keeping the total system bounded. As in section \ref{sect3}, we require the following for 
the amplitude change in the last order,
\begin{equation} 
|\theta_{2}|^{2} \ < \ |\theta_{0}||\theta_{4}|
\end{equation} 
The above inequality must hold if the amplitude is to flip signs. The change of signs in the amplitude is essential for consistency; however, there seems to be no 
deep reason why the inequality should always hold. \\
As mentioned in the beginning, the time scale for which the approximate solution in section \ref{sect3} is valid will increase as the forcing term amplitude 
decreases. Fig. \ref{figdiff} shows two other cases with smaller values of $F$; it is evident that the timescale of validity has increased significantly. One should 
also note that our approach in section \ref{sect3} is different to the standard approach when dealing with nonlinear oscillators like the forced Duffing oscillator 
(see \cite{duff}). We have straightaway considered a perturbation in $\theta$ with the forcing amplitude acting as the regulator without considering the frequency 
of the oscillator or the forcing term in the expansion (no additional frequency - amplitude constraints). \\
We have presented two interesting results from the more conventional approaches - one, the maximum amplitude of the response scales as $F^{1/3}$, and two, the time 
period of the slower modulation period scales as $F^{-2/3}$, where $F$ is the dimensionless drive introduced in section \ref{sect3}. Similar beat behaviour has been 
experimentally observed in a related system in \cite{Cross}. We have also briefly considered the system at larger drive values - our numerical investigation is 
presented in section \ref{sect4}. There is an interesting first-order transitional behaviour which is $F$-dependent with the maximum Lyapunov exponent showing a 
sudden jump at around $F \approx 0.33$ - this is evident from both the plots depicted in Fig. \ref{figlya}. The stroboscopic maps and phase space diagrams in 
Fig. \ref{figPM} also show a break in periodicity associated with the $F$ value. Taken together, the sinusoidal oscillator presents a rich variety of features.

\section{Acknowledgements}
The computational work was carried out using the SciML ecosystem on Julia (\cite{Datseris2018}, \cite{ma2021modelingtoolkit}, 
\cite{rackauckas2017differentialequations}). 
This work was entirely funded by the authors' respective institutes. SH would like to acknowledge his KVPY fellowship provided by the DST and enlightening discussions
with the Julia community members on the Julia forum. 
\subsection{Author contribution statement}
Both the authors contributed equally to setting up the problem. Bulk of the calculation and numerics was done by SH.
\subsection{Data availability statement}
The code used for the numerical results is publicly available in a GitHub repository 
\href{https://github.com/PsiHQ/2305.04125_ForcedOscillator}{\textcolor{blue}{here}}. 

\bibliography{Bibliography}

\providecommand{\noopsort}[1]{}\providecommand{\singleletter}[1]{#1}%
\begin{thebibliography}{23}%
\makeatletter
\providecommand \@ifxundefined [1]{%
 \@ifx{#1\undefined}
}%
\providecommand \@ifnum [1]{%
 \ifnum #1\expandafter \@firstoftwo
 \else \expandafter \@secondoftwo
 \fi
}%
\providecommand \@ifx [1]{%
 \ifx #1\expandafter \@firstoftwo
 \else \expandafter \@secondoftwo
 \fi
}%
\providecommand \natexlab [1]{#1}%
\providecommand \enquote  [1]{``#1''}%
\providecommand \bibnamefont  [1]{#1}%
\providecommand \bibfnamefont [1]{#1}%
\providecommand \citenamefont [1]{#1}%
\providecommand \href@noop [0]{\@secondoftwo}%
\providecommand \href [0]{\begingroup \@sanitize@url \@href}%
\providecommand \@href[1]{\@@startlink{#1}\@@href}%
\providecommand \@@href[1]{\endgroup#1\@@endlink}%
\providecommand \@sanitize@url [0]{\catcode `\\12\catcode `\$12\catcode
  `\&12\catcode `\#12\catcode `\^12\catcode `\_12\catcode `\%12\relax}%
\providecommand \@@startlink[1]{}%
\providecommand \@@endlink[0]{}%
\providecommand \url  [0]{\begingroup\@sanitize@url \@url }%
\providecommand \@url [1]{\endgroup\@href {#1}{\urlprefix }}%
\providecommand \urlprefix  [0]{URL }%
\providecommand \Eprint [0]{\href }%
\providecommand \doibase [0]{https://doi.org/}%
\providecommand \selectlanguage [0]{\@gobble}%
\providecommand \bibinfo  [0]{\@secondoftwo}%
\providecommand \bibfield  [0]{\@secondoftwo}%
\providecommand \translation [1]{[#1]}%
\providecommand \BibitemOpen [0]{}%
\providecommand \bibitemStop [0]{}%
\providecommand \bibitemNoStop [0]{.\EOS\space}%
\providecommand \EOS [0]{\spacefactor3000\relax}%
\providecommand \BibitemShut  [1]{\csname bibitem#1\endcsname}%
\let\auto@bib@innerbib\@empty
\bibitem [{\citenamefont {Aiello}\ \emph {et~al.}(2019)\citenamefont {Aiello},
  \citenamefont {Harris},\ and\ \citenamefont {Marquardt}}]{aiello}%
  \BibitemOpen
  \bibfield  {author} {\bibinfo {author} {\bibfnamefont {A.}~\bibnamefont
  {Aiello}}, \bibinfo {author} {\bibfnamefont {J.~G.~E.}\ \bibnamefont
  {Harris}},\ and\ \bibinfo {author} {\bibfnamefont {F.}~\bibnamefont
  {Marquardt}},\ }\bibfield  {title} {\bibinfo {title} {Perturbation theory of
  optical resonances of deformed dielectric spheres},\ }\href
  {https://doi.org/10.1103/PhysRevA.100.023837} {\bibfield  {journal} {\bibinfo
   {journal} {Phys. Rev. A}\ }\textbf {\bibinfo {volume} {100}},\ \bibinfo
  {pages} {023837} (\bibinfo {year} {2019})}\BibitemShut {NoStop}%
\bibitem [{\citenamefont {Gohsrich}\ \emph {et~al.}(2021)\citenamefont
  {Gohsrich}, \citenamefont {Shah},\ and\ \citenamefont {Aiello}}]{aiello2}%
  \BibitemOpen
  \bibfield  {author} {\bibinfo {author} {\bibfnamefont {J.}~\bibnamefont
  {Gohsrich}}, \bibinfo {author} {\bibfnamefont {T.}~\bibnamefont {Shah}},\
  and\ \bibinfo {author} {\bibfnamefont {A.}~\bibnamefont {Aiello}},\
  }\bibfield  {title} {\bibinfo {title} {Perturbation theory of nearly
  spherical dielectric optical resonators},\ }\href
  {https://doi.org/10.1103/PhysRevA.104.023516} {\bibfield  {journal} {\bibinfo
   {journal} {Phys. Rev. A}\ }\textbf {\bibinfo {volume} {104}},\ \bibinfo
  {pages} {023516} (\bibinfo {year} {2021})}\BibitemShut {NoStop}%
\bibitem [{\citenamefont {Foreman}\ \emph {et~al.}(2015)\citenamefont
  {Foreman}, \citenamefont {Swaim},\ and\ \citenamefont {Vollmer}}]{foreman}%
  \BibitemOpen
  \bibfield  {author} {\bibinfo {author} {\bibfnamefont {M.~R.}\ \bibnamefont
  {Foreman}}, \bibinfo {author} {\bibfnamefont {J.~D.}\ \bibnamefont {Swaim}},\
  and\ \bibinfo {author} {\bibfnamefont {F.}~\bibnamefont {Vollmer}},\
  }\bibfield  {title} {\bibinfo {title} {Whispering gallery mode sensors},\
  }\href {https://doi.org/10.1364/AOP.7.000168} {\bibfield  {journal} {\bibinfo
   {journal} {Adv. Opt. Photon.}\ }\textbf {\bibinfo {volume} {7}},\ \bibinfo
  {pages} {168} (\bibinfo {year} {2015})}\BibitemShut {NoStop}%
\bibitem [{\citenamefont {Childress}\ \emph {et~al.}(2017)\citenamefont
  {Childress}, \citenamefont {Schmidt}, \citenamefont {Kashkanova},
  \citenamefont {Brown}, \citenamefont {Harris}, \citenamefont {Aiello},
  \citenamefont {Marquardt},\ and\ \citenamefont {Harris}}]{childress}%
  \BibitemOpen
  \bibfield  {author} {\bibinfo {author} {\bibfnamefont {L.}~\bibnamefont
  {Childress}}, \bibinfo {author} {\bibfnamefont {M.~P.}\ \bibnamefont
  {Schmidt}}, \bibinfo {author} {\bibfnamefont {A.~D.}\ \bibnamefont
  {Kashkanova}}, \bibinfo {author} {\bibfnamefont {C.~D.}\ \bibnamefont
  {Brown}}, \bibinfo {author} {\bibfnamefont {G.~I.}\ \bibnamefont {Harris}},
  \bibinfo {author} {\bibfnamefont {A.}~\bibnamefont {Aiello}}, \bibinfo
  {author} {\bibfnamefont {F.}~\bibnamefont {Marquardt}},\ and\ \bibinfo
  {author} {\bibfnamefont {J.~G.~E.}\ \bibnamefont {Harris}},\ }\bibfield
  {title} {\bibinfo {title} {Cavity optomechanics in a levitated helium drop},\
  }\href {https://doi.org/10.1103/PhysRevA.96.063842} {\bibfield  {journal}
  {\bibinfo  {journal} {Phys. Rev. A}\ }\textbf {\bibinfo {volume} {96}},\
  \bibinfo {pages} {063842} (\bibinfo {year} {2017})}\BibitemShut {NoStop}%
\bibitem [{\citenamefont {Elliott}(1982)}]{elliot}%
  \BibitemOpen
  \bibfield  {author} {\bibinfo {author} {\bibfnamefont {J.~A.}\ \bibnamefont
  {Elliott}},\ }\bibfield  {title} {\bibinfo {title} {{Nonlinear resonance in
  vibrating strings}},\ }\href {https://doi.org/10.1119/1.12896} {\bibfield
  {journal} {\bibinfo  {journal} {American Journal of Physics}\ }\textbf
  {\bibinfo {volume} {50}},\ \bibinfo {pages} {1148} (\bibinfo {year}
  {1982})}\BibitemShut {NoStop}%
\bibitem [{\citenamefont {Mawhin}(1988)}]{Jean}%
  \BibitemOpen
  \bibfield  {author} {\bibinfo {author} {\bibfnamefont {J.}~\bibnamefont
  {Mawhin}},\ }\bibfield  {title} {\bibinfo {title} {The forced pendulum: A
  paradigm for nonlinear analysis and dynamical systems},\ }\href@noop {}
  {\bibfield  {journal} {\bibinfo  {journal} {Expo. Math.}\ }\textbf {\bibinfo
  {volume} {6.3}},\ \bibinfo {pages} {271} (\bibinfo {year}
  {1988})}\BibitemShut {NoStop}%
\bibitem [{\citenamefont {Grebogi}\ \emph {et~al.}(1987)\citenamefont
  {Grebogi}, \citenamefont {Ott}, \citenamefont {Romeiras},\ and\ \citenamefont
  {Yorke}}]{Yorke}%
  \BibitemOpen
  \bibfield  {author} {\bibinfo {author} {\bibfnamefont {C.}~\bibnamefont
  {Grebogi}}, \bibinfo {author} {\bibfnamefont {E.}~\bibnamefont {Ott}},
  \bibinfo {author} {\bibfnamefont {F.}~\bibnamefont {Romeiras}},\ and\
  \bibinfo {author} {\bibfnamefont {J.~A.}\ \bibnamefont {Yorke}},\ }\bibfield
  {title} {\bibinfo {title} {Critical exponents for crisis-induced
  intermittency},\ }\href {https://doi.org/10.1103/PhysRevA.36.5365} {\bibfield
   {journal} {\bibinfo  {journal} {Phys. Rev. A}\ }\textbf {\bibinfo {volume}
  {36}},\ \bibinfo {pages} {5365} (\bibinfo {year} {1987})}\BibitemShut
  {NoStop}%
\bibitem [{\citenamefont {Harish}\ \emph {et~al.}(2002)\citenamefont {Harish},
  \citenamefont {Rajasekar},\ and\ \citenamefont {Murthy}}]{harish}%
  \BibitemOpen
  \bibfield  {author} {\bibinfo {author} {\bibfnamefont {R.}~\bibnamefont
  {Harish}}, \bibinfo {author} {\bibfnamefont {S.}~\bibnamefont {Rajasekar}},\
  and\ \bibinfo {author} {\bibfnamefont {K.~P.~N.}\ \bibnamefont {Murthy}},\
  }\bibfield  {title} {\bibinfo {title} {Diffusion in a periodically driven
  damped and undamped pendulum},\ }\href
  {https://doi.org/10.1103/PhysRevE.65.046214} {\bibfield  {journal} {\bibinfo
  {journal} {Phys. Rev. E}\ }\textbf {\bibinfo {volume} {65}},\ \bibinfo
  {pages} {046214} (\bibinfo {year} {2002})}\BibitemShut {NoStop}%
\bibitem [{\citenamefont {Gandhi}\ \emph {et~al.}(2015)\citenamefont {Gandhi},
  \citenamefont {Knobloch},\ and\ \citenamefont {Beaume}}]{Adler}%
  \BibitemOpen
  \bibfield  {author} {\bibinfo {author} {\bibfnamefont {P.}~\bibnamefont
  {Gandhi}}, \bibinfo {author} {\bibfnamefont {E.}~\bibnamefont {Knobloch}},\
  and\ \bibinfo {author} {\bibfnamefont {C.}~\bibnamefont {Beaume}},\
  }\bibfield  {title} {\bibinfo {title} {Dynamics of phase slips in systems
  with time-periodic modulation},\ }\href
  {https://doi.org/10.1103/PhysRevE.92.062914} {\bibfield  {journal} {\bibinfo
  {journal} {Phys. Rev. E}\ }\textbf {\bibinfo {volume} {92}},\ \bibinfo
  {pages} {062914} (\bibinfo {year} {2015})}\BibitemShut {NoStop}%
\bibitem [{\citenamefont {Lazer}\ and\ \citenamefont {Leach}(1969)}]{lazer}%
  \BibitemOpen
  \bibfield  {author} {\bibinfo {author} {\bibfnamefont {A.~C.}\ \bibnamefont
  {Lazer}}\ and\ \bibinfo {author} {\bibfnamefont {D.~E.}\ \bibnamefont
  {Leach}},\ }\bibfield  {title} {\bibinfo {title} {Bounded perturbations of
  forced harmonic oscillators at resonance},\ }\href@noop {} {\bibfield
  {journal} {\bibinfo  {journal} {Annali di Matematica Pura ed Applicata}\
  }\textbf {\bibinfo {volume} {82}},\ \bibinfo {pages} {49} (\bibinfo {year}
  {1969})}\BibitemShut {NoStop}%
\bibitem [{\citenamefont {Alonso}\ and\ \citenamefont {Ortega}(1996)}]{alonso}%
  \BibitemOpen
  \bibfield  {author} {\bibinfo {author} {\bibfnamefont {J.}~\bibnamefont
  {Alonso}}\ and\ \bibinfo {author} {\bibfnamefont {R.}~\bibnamefont
  {Ortega}},\ }\bibfield  {title} {\bibinfo {title} {Unbounded solutions of
  semilinear equations at resonance},\ }\href
  {https://doi.org/10.1088/0951-7715/9/5/003} {\bibfield  {journal} {\bibinfo
  {journal} {Nonlinearity}\ }\textbf {\bibinfo {volume} {9}},\ \bibinfo {pages}
  {1099} (\bibinfo {year} {1996})}\BibitemShut {NoStop}%
\bibitem [{\citenamefont {Lai}(1984)}]{Lai}%
  \BibitemOpen
  \bibfield  {author} {\bibinfo {author} {\bibfnamefont {H.~M.}\ \bibnamefont
  {Lai}},\ }\bibfield  {title} {\bibinfo {title} {{On the recurrence phenomenon
  of a resonant spring pendulum}},\ }\href {https://doi.org/10.1119/1.13696}
  {\bibfield  {journal} {\bibinfo  {journal} {American Journal of Physics}\
  }\textbf {\bibinfo {volume} {52}},\ \bibinfo {pages} {219} (\bibinfo {year}
  {1984})}\BibitemShut {NoStop}%
\bibitem [{\citenamefont {Olsson}(1976)}]{olsson}%
  \BibitemOpen
  \bibfield  {author} {\bibinfo {author} {\bibfnamefont {M.~G.}\ \bibnamefont
  {Olsson}},\ }\bibfield  {title} {\bibinfo {title} {Why does a mass on a
  spring sometimes misbehave},\ }\href@noop {} {\bibfield  {journal} {\bibinfo
  {journal} {American Journal of Physics}\ }\textbf {\bibinfo {volume} {44}},\
  \bibinfo {pages} {1211} (\bibinfo {year} {1976})}\BibitemShut {NoStop}%
\bibitem [{\citenamefont {Hohenberg}\ and\ \citenamefont
  {Halperin}(1977)}]{halp}%
  \BibitemOpen
  \bibfield  {author} {\bibinfo {author} {\bibfnamefont {P.~C.}\ \bibnamefont
  {Hohenberg}}\ and\ \bibinfo {author} {\bibfnamefont {B.~I.}\ \bibnamefont
  {Halperin}},\ }\bibfield  {title} {\bibinfo {title} {Theory of dynamic
  critical phenomena},\ }\href {https://doi.org/10.1103/RevModPhys.49.435}
  {\bibfield  {journal} {\bibinfo  {journal} {Rev. Mod. Phys.}\ }\textbf
  {\bibinfo {volume} {49}},\ \bibinfo {pages} {435} (\bibinfo {year}
  {1977})}\BibitemShut {NoStop}%
\bibitem [{\citenamefont {Benettin}\ \emph {et~al.}(1976)\citenamefont
  {Benettin}, \citenamefont {Galgani},\ and\ \citenamefont
  {Strelcyn}}]{Bennetin}%
  \BibitemOpen
  \bibfield  {author} {\bibinfo {author} {\bibfnamefont {G.}~\bibnamefont
  {Benettin}}, \bibinfo {author} {\bibfnamefont {L.}~\bibnamefont {Galgani}},\
  and\ \bibinfo {author} {\bibfnamefont {J.-M.}\ \bibnamefont {Strelcyn}},\
  }\bibfield  {title} {\bibinfo {title} {Kolmogorov entropy and numerical
  experiments},\ }\href {https://doi.org/10.1103/PhysRevA.14.2338} {\bibfield
  {journal} {\bibinfo  {journal} {Phys. Rev. A}\ }\textbf {\bibinfo {volume}
  {14}},\ \bibinfo {pages} {2338} (\bibinfo {year} {1976})}\BibitemShut
  {NoStop}%
\bibitem [{\citenamefont {Balcerzak}\ \emph {et~al.}(2018)\citenamefont
  {Balcerzak}, \citenamefont {Pikunov},\ and\ \citenamefont
  {Dabrowski}}]{balcerzak}%
  \BibitemOpen
  \bibfield  {author} {\bibinfo {author} {\bibfnamefont {M.}~\bibnamefont
  {Balcerzak}}, \bibinfo {author} {\bibfnamefont {D.}~\bibnamefont {Pikunov}},\
  and\ \bibinfo {author} {\bibfnamefont {A.}~\bibnamefont {Dabrowski}},\
  }\bibfield  {title} {\bibinfo {title} {The fastest, simplified method of
  lyapunov exponents spectrum estimation for continuous-time dynamical
  systems},\ }\href {https://doi.org/https://doi.org/10.1007/s11071-018-4544-z}
  {\bibfield  {journal} {\bibinfo  {journal} {Nonlinear Dyn}\ }\textbf
  {\bibinfo {volume} {94}},\ \bibinfo {pages} {3053–3065} (\bibinfo {year}
  {2018})}\BibitemShut {NoStop}%
\bibitem [{\citenamefont {Stoop}\ and\ \citenamefont {Parisi}(1993)}]{Stoop}%
  \BibitemOpen
  \bibfield  {author} {\bibinfo {author} {\bibfnamefont {R.}~\bibnamefont
  {Stoop}}\ and\ \bibinfo {author} {\bibfnamefont {J.}~\bibnamefont {Parisi}},\
  }\bibfield  {title} {\bibinfo {title} {On the scaling function of lyapunov
  exponents for intermittent maps},\ }\href
  {https://doi.org/doi:10.1515/zna-1993-5-609} {\bibfield  {journal} {\bibinfo
  {journal} {Zeitschrift für Naturforschung A}\ }\textbf {\bibinfo {volume}
  {48}},\ \bibinfo {pages} {641} (\bibinfo {year} {1993})}\BibitemShut
  {NoStop}%
\bibitem [{\citenamefont {Kalm{\'a}r-Nagy}\ and\ \citenamefont
  {Balachandran}(2011)}]{duff}%
  \BibitemOpen
  \bibfield  {author} {\bibinfo {author} {\bibfnamefont {T.}~\bibnamefont
  {Kalm{\'a}r-Nagy}}\ and\ \bibinfo {author} {\bibfnamefont {B.}~\bibnamefont
  {Balachandran}},\ }\bibfield  {title} {\bibinfo {title} {Forced harmonic
  vibration of a duffing oscillator with linear viscous damping},\ }in\
  \href@noop {} {\emph {\bibinfo {booktitle} {The duffing equation: nonlinear
  oscillators and their behaviour}}}\ (\bibinfo {year} {2011})\BibitemShut
  {NoStop}%
\bibitem [{\citenamefont {Cross}(2021)}]{Cross}%
  \BibitemOpen
  \bibfield  {author} {\bibinfo {author} {\bibfnamefont {R.}~\bibnamefont
  {Cross}},\ }\bibfield  {title} {\bibinfo {title} {Observations of a driven
  pendulum at low amplitudes},\ }\href
  {https://doi.org/10.1088/1361-6404/ac059b} {\bibfield  {journal} {\bibinfo
  {journal} {European Journal of Physics}\ }\textbf {\bibinfo {volume} {42}},\
  \bibinfo {pages} {055001} (\bibinfo {year} {2021})}\BibitemShut {NoStop}%
\bibitem [{\citenamefont {Datseris}(2018)}]{Datseris2018}%
  \BibitemOpen
  \bibfield  {author} {\bibinfo {author} {\bibfnamefont {G.}~\bibnamefont
  {Datseris}},\ }\bibfield  {title} {\bibinfo {title} {Dynamicalsystems.jl: A
  julia software library for chaos and nonlinear dynamics},\ }\href
  {https://doi.org/10.21105/joss.00598} {\bibfield  {journal} {\bibinfo
  {journal} {Journal of Open Source Software}\ }\textbf {\bibinfo {volume}
  {3}},\ \bibinfo {pages} {598} (\bibinfo {year} {2018})}\BibitemShut {NoStop}%
\bibitem [{\citenamefont {Ma}\ \emph {et~al.}(2021)\citenamefont {Ma},
  \citenamefont {Gowda}, \citenamefont {Anantharaman}, \citenamefont
  {Laughman}, \citenamefont {Shah},\ and\ \citenamefont
  {Rackauckas}}]{ma2021modelingtoolkit}%
  \BibitemOpen
  \bibfield  {author} {\bibinfo {author} {\bibfnamefont {Y.}~\bibnamefont
  {Ma}}, \bibinfo {author} {\bibfnamefont {S.}~\bibnamefont {Gowda}}, \bibinfo
  {author} {\bibfnamefont {R.}~\bibnamefont {Anantharaman}}, \bibinfo {author}
  {\bibfnamefont {C.}~\bibnamefont {Laughman}}, \bibinfo {author}
  {\bibfnamefont {V.}~\bibnamefont {Shah}},\ and\ \bibinfo {author}
  {\bibfnamefont {C.}~\bibnamefont {Rackauckas}},\ }\href@noop {} {\bibinfo
  {title} {Modelingtoolkit: A composable graph transformation system for
  equation-based modeling}} (\bibinfo {year} {2021}),\ \Eprint
  {https://arxiv.org/abs/2103.05244} {arXiv:2103.05244 [cs.MS]} \BibitemShut
  {NoStop}%
\bibitem [{\citenamefont {Rackauckas}\ and\ \citenamefont
  {Nie}(2017)}]{rackauckas2017differentialequations}%
  \BibitemOpen
  \bibfield  {author} {\bibinfo {author} {\bibfnamefont {C.}~\bibnamefont
  {Rackauckas}}\ and\ \bibinfo {author} {\bibfnamefont {Q.}~\bibnamefont
  {Nie}},\ }\bibfield  {title} {\bibinfo {title} {Differential{E}quations.jl--a
  performant and feature-rich ecosystem for solving differential equations in
  {J}ulia},\ }\href@noop {} {\bibfield  {journal} {\bibinfo  {journal} {Journal
  of Open Research Software}\ }\textbf {\bibinfo {volume} {5}} (\bibinfo {year}
  {2017})}\BibitemShut {NoStop}%
\bibitem [{\citenamefont {Guo}\ and\ \citenamefont {Cao}(2014)}]{Guo}%
  \BibitemOpen
  \bibfield  {author} {\bibinfo {author} {\bibfnamefont {K.}~\bibnamefont
  {Guo}}\ and\ \bibinfo {author} {\bibfnamefont {S.}~\bibnamefont {Cao}},\
  }\bibfield  {title} {\bibinfo {title} {Modified lindstedt-poincaré method
  for obtaining resonance periodic solutions of nonlinear non-autonomous
  oscillators},\ }\bibfield  {journal} {\bibinfo  {journal} {Trans. Tianjin
  Univ.}\ }\textbf {\bibinfo {volume} {20}},\ \href
  {https://doi.org/10.1007/s12209-014-2126-9} {10.1007/s12209-014-2126-9}
  (\bibinfo {year} {2014})\BibitemShut {NoStop}%
\end{thebibliography}%


\providecommand{\noopsort}[1]{}\providecommand{\singleletter}[1]{#1}%
%
\nocite{*}
\end{document}